\documentclass[apj]{emulateapj}

\usepackage{longtable}
\usepackage{comment}
\usepackage{color}
\usepackage[normalem]{ulem}

\usepackage[colorlinks=true, pdfstartview=FitV, linkcolor=black,citecolor=black, urlcolor=blue]{hyperref}

\newcommand{\cxc}{{\em Chandra}}
\newcommand{\xmm}{XMM-{\em Newton}}

\newcommand{\rcn}{118 $\pm$ 9}
\newcommand{\betan}{0.54 $\pm$ 0.02}

\newcommand{\rtn}{1266 $\pm$ 243}

\newcommand{\Tn}{2.74 $\pm$ 0.14}

\newcommand{\rcs}{228 $\pm$ 12}
\newcommand{\betas}{0.48 $\pm$ 0.01}

\newcommand{\rts}{1459 $\pm$ 141}

\newcommand{\Ts}{4.07 $\pm$ 0.22}
 
\newcommand{\chiba}{4.952}
\newcommand{\alphaba}{75.13}
\newcommand{\Rba}{1559.5}
\newcommand{\Rmba}{4089.6}
\newcommand{\Vba}{1173.1}

\newcommand{\chibb}{5.600}
\newcommand{\alphabb}{25.48}
\newcommand{\Rbb}{443.1}
\newcommand{\Rmbb}{3952.9}
\newcommand{\Vbb}{2635.6}

\newcommand{\cchi}{3.157}
\newcommand{\aalpha}{88.67}
\newcommand{\R}{17213.7}
\newcommand{\V}{1134.1}
\newcommand{\Vinf}{1041.5}

\newcommand{\rhydn}{$724 \pm 23$}
\newcommand{\kThydn}{$2.11 \pm 0.16$}
\newcommand{\ghydn}{$(1.45 \pm 0.23) \times 10^{13}$}
\newcommand{\mhydn}{$(1.13 \pm 0.11) \times 10^{14}$}
\newcommand{\fghydn}{$0.130 \pm 0.023$}

\newcommand{\rhyds}{$836 \pm 25$}
\newcommand{\kThyds}{$2.85 \pm 0.20$}
\newcommand{\ghyds}{$(1.45 \pm 0.14) \times 10^{13}$}
\newcommand{\mhyds}{$(1.73 \pm 0.15) \times 10^{14}$}
\newcommand{\fghyds}{$0.084 \pm 0.008$}

\newcommand{\ryxn}{$771 \pm 47$}
\newcommand{\kTyxn}{$2.08 \pm 0.17$}
\newcommand{\gyxn}{$(1.62 \pm 0.38) \times 10^{13}$}
\newcommand{\myxn}{$(1.37 \pm 0.25) \times 10^{14}$}
\newcommand{\fgyxn}{$0.117 \pm 0.015$}

\newcommand{\ryxs}{$793 \pm 42$}
\newcommand{\kTyxs}{$2.90 \pm 0.21$}
\newcommand{\gyxs}{$(1.33 \pm 0.21) \times 10^{13}$}
\newcommand{\myxs}{$(1.48 \pm 0.24) \times 10^{14}$}
\newcommand{\fgyxs}{$0.090 \pm 0.007$}

\newcommand{\nOn}{$(3.19 \pm 0.28) \times 10^{-3}$}
\newcommand{\rhogn}{$(6.21 \pm 0.54) \times 10^{-27}$}

\newcommand{\nOs}{$(8.96 \pm 0.51) \times 10^{-4}$}
\newcommand{\rhogs}{$(1.74 \pm 0.10) \times 10^{-27}$}

\shorttitle{PLCKG345.40-39.34: a {\it Planck} SZ-detected Bimodal Cluster}
\shortauthors{Andrade-Santos et al.}

\begin{document}


\title{{\em Chandra} and XMM-{\em Newton} Observations of the Bimodal
  {\em Planck} SZ-detected Cluster PLCKG345.40-39.34 (A3716) with High and Low Entropy Subcluster Cores}


\author{
Felipe Andrade-Santos$^1$, 
Christine Jones$^1$,  
William R. Forman$^1$,
Stephen S. Murray$^{1,2}$,
Ralph P. Kraft$^1$,
Alexey Vikhlinin$^1$, 
Reinout J. van Weeren$^1$, 
Paul E. J. Nulsen$^1$,
Laurence P. David$^1$,
William A. Dawson$^3$,
Monique Arnaud$^4$,
Etienne Pointecouteau$^{5,6}$,
Gabriel W. Pratt$^4$, \&
Jean-Baptiste Melin$^7$}
\affil{$^1$Harvard-Smithsonian Center for Astrophysics, 60 Garden Street, Cambridge, MA 02138, USA \\
$^2$Department of Physics and Astronomy, The Johns Hopkins University,
  3400 N. Charles St., Baltimore, MD 21218, USA  \\ 
$^3$Lawrence Livermore National Laboratory, P.O. Box 808 L-
210, Livermore, CA, 94551, USA \\
$^4$Laboratoire AIM, IRFU/Service d'Astrophysique – CEA/DSM – CNRS – Universit\'e Paris Diderot, B\^at. 709, CEA-Saclay,
91191 Gif-sur-Yvette Cedex, France \\
$^5$Universit\'e de Toulouse, UPS-OMP, IRAP, F-31028 Toulouse Cedex
4, France \\
$^6$CNRS, IRAP, 9 Av. colonel Roche, BP 44346, F-31028 Toulouse
Cedex 4, France \\
$^7$DSM/Irfu/SPP, CEA-Saclay, F-91191 Gif-sur-Yvette Cedex, France}




\begin{abstract}

We present results from {\em Chandra}, XMM-{\em Newton}, and ROSAT
observations of the {\em Planck} SZ-detected cluster A3716
(PLCKG345.40-39.34 -- G345). We show that G345 is, in fact, 
two subclusters separated on the sky by 400 kpc. We measure the
subclusters' gas temperatures ($\sim$ 2--3 keV), total ($\sim$ 1--2 $\times 10^{14}  
~M_\odot$) and gas ($\sim$ 1--2 $\times 10^{13} ~M_\odot$) masses,
gas mass fraction within $r_{500}$, entropy profiles, and X-ray luminosities
($\sim 10^{43} \rm ~erg~s^{-1}$).  Using the gas density and temperature profiles
for both subclusters, we show that there is good ($0.8\sigma$) agreement
between the expected Sunyaev-Zel'dovich signal predicted from the X-ray data and
that measured   from the {\em Planck} mission, and better agreement within
$0.6\sigma$ when we re-computed the 
{\em Planck} value assuming a two component cluster model, with
relative amplitudes  fixed based on the X-ray data. Dynamical analysis
shows that the two galaxy subclusters are very likely ($> 97\%$ probability)
gravitationally bound, and in the most likely scenario, the subclusters will undergo core
passage in $500 \pm 200$ Myr. The northern subcluster is centrally peaked
and has a low  entropy core, while the southern subcluster has a high central
entropy. The high central entropy in the southern subcluster can be explained
either by the mergers of several groups, as suggested by the presence of
five giant ellipticals or by AGN energy injection, as suggested by the
presence  of a strong radio source in one of its massive elliptical
galaxies, or by a combination of both processes.

\end{abstract}



\section{Introduction}

Clusters of galaxies are the largest gravitationally bound structures in the Universe
to have reached virial equilibrium. In the standard $\Lambda$CDM cosmology,
massive halos dominated by
dark matter assemble by accretion of smaller groups and clusters. 
Under the influence of gravity, uncollapsed matter and smaller collapsed halos fall into
larger halos and, occasionally, halos of comparable mass merge with
one another. Observations of substructures in clusters of galaxies 
\citep[e.g.,][]{1984Jones,1999Jones,1995Mohr,1996Buote,2005Jeltema,2010Lagana,
2012Andrade-Santos,2013Andrade-Santos} and the growth of structure \citep{2009bVik}
strongly suggest that clusters formed recently \citep{1982Forman,1992Richstone}.
To better understand the evolution of the Universe, it is important to 
identify and characterize these large structures.

The first catalog of 189 Sunyaev-Zel'dovich (SZ) clusters detected by
the {\em Planck} mission was released
in early 2011 \citep{2011PlanckCol}.
Through a \cxc ~XVP (X-ray Visionary Program -- PI: Christine Jones) and HRC
Guaranteed Time Observations (PI: Stephen S. Murray), which comprise the {\em
Chandra-Planck} Legacy Program for Massive Clusters of 
Galaxies\footnote{\scriptsize \href{http://hea-www.cfa.harvard.edu/CHANDRA_PLANCK_CLUSTERS/}{hea-www.cfa.harvard.edu/CHANDRA\_PLANCK\_CLUSTERS/}},
we are obtaining 
\cxc ~exposures sufficient to collect at least 10,000 source counts for each of the ESZ
{\em Planck} clusters at $z \le 0.35$. 
PLCKG345.40-39.34 (hereafter G345 - RA = 20:52:16.8, DEC = -52:49:30.7) is a nearby cluster (z = 0.045). 
G345 was the first cluster observed as part of the {\em Chandra} XVP, which
revealed it to be a double cluster (we distinguish the north and south
subclusters as G345N
and G345S, respectively). 

\begin{figure*}[hbt!]
\centerline{%
\includegraphics[width=1.05\textwidth]{%
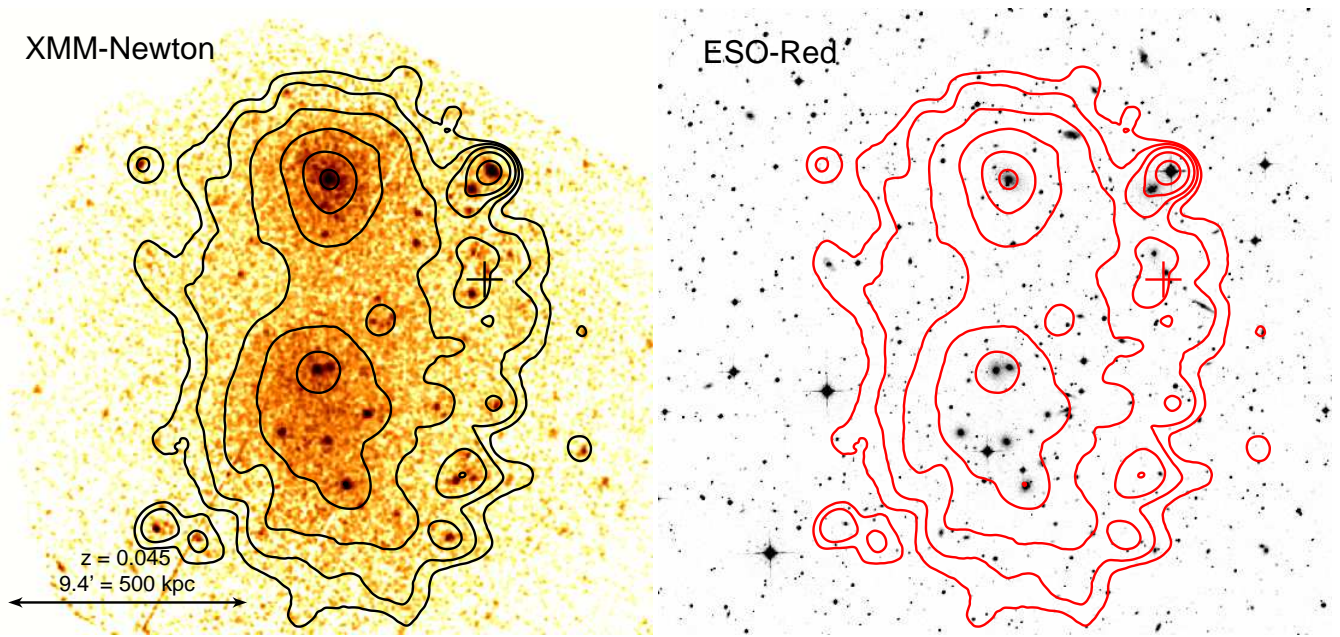}}
\caption{
\small{
Left: 0.5--2.0 keV XMM-{\em Newton} image of G345. Right: ESO DSS-Red optical
image of the same sky region presented in the left panel. Isointensity
X-ray contours are shown in both panels. The cross indicates the
\citet{1989Abell} catalogue position for A3716. \label{fig:xmm_esored} 
}
}
\end{figure*}

Historically, \citet{1989Abell} had catalogued a single cluster of
galaxies (A3716 -- cross mark in Figure \ref{fig:xmm_esored})
located $\sim 6'$ northwest of G345S and $\sim 6'$
southwest of G345N. 

In Section 2 of this paper we present the {\em Chandra}, XMM-{\em
  Newton}, and ROSAT observations and data reduction. In Section 3 we describe 
the X-ray spatial and spectral analysis. The total mass and gas mass for each
subcluster are presented in Section 4, 
and the entropy profiles of both subclusters are computed in Section
5. In Section 6 we compute the
expected Sunyaev-Zel'dovich signal from the X-ray
data, as well as a comparison between the X-ray emission and the {\em
  Planck} reconstructed Sunyaev-Zel'dovich map. Finally, we present a dynamical
model for the two subclusters in Section 7 and the conclusions in Section 8. 

We assume a standard $\Lambda$CDM cosmology with $\Omega_{\rm M}=0.3$,
$\Omega_{\Lambda}=0.7$ and $H_0=70$~km~s$^{-1}$Mpc$^{-1}$, implying a
linear scale of $0.89\rm\ kpc\ arcsec^{-1}$ at the G345 luminosity distance of
199 Mpc ($z=0.045$).

\begin{deluxetable*}{ccccc}
\tablecaption{Best Fit Parameters for the $\beta$-Model
   and central density.} 
\tablewidth{0pt} 
\tablehead{ 
\colhead{Subcluster}&
\colhead{$r_{\rm c}$} &
\colhead{$\beta$} &
\colhead{$n_{\rm e,0}$} &
\colhead{$\rho_{0}$}  \\
\colhead{}&
\colhead{(kpc)} &
\colhead{} &
\colhead{($\rm cm^{-3}$)} &
\colhead{($\rm g ~cm^{-3}$)}
}
\startdata 
G345N & $ $\rcn$ $ & $ $\betan$ $ & $ $\nOn$ $ & $ $\rhogn$ $\\
G345S & $ $\rcs$ $ & $ $\betas$ $ & $ $\nOs$ $ & $ $\rhogs$ $
\enddata
\tablecomments{Columns list subclusters' names, determination and 1$\sigma$ uncertainties for the core radius,
  $\beta$ (Equation (\ref{BS_beta})), and central electron number and
  gas mass densities (Equation (\ref{n0})).}
\label{tab:beta_model}
\end{deluxetable*}


\begin{deluxetable*}{ccccc}
\tablecaption{Comparison Between the Best Fit Parameters for the
  $\beta$-Model Using Different Data Sets.} 
\tablewidth{0pt} 
\tablehead{ 
\colhead{Telescope}&
\colhead{Subcluster}&
\colhead{$r_{\rm c}$} &
\colhead{$\beta$} &
\colhead{$\chi^2_{\rm red}$}  \\
\colhead{}&
\colhead{}&
\colhead{(kpc)} &
\colhead{} &
\colhead{}
}
\startdata 
{\em Chandra} & G345N & $ 82 \pm 23 $ & $ 0.44 \pm 0.08 $ & 0.83 \\
    & G345S & $ 228 \pm 49$ & $ 0.44 \pm 0.04 $ & 1.37 \\
XMM-{\em Newton} & G345N & $ 91 \pm 33 $  & $ 0.46 \pm 0.10 $ & 1.05 \\
    & G345S & $ 199 \pm 20 $ & $ 0.45 \pm 0.02 $ & 1.73 \\
{\em Chandra} + XMM-{\em Newton} + ROSAT & G345N & $ $\rcn$ $ & $ $\betan$ $ & 1.07 \\
    & G345S & $ $\rcs$ $ & $ $\betas$ $ & 1.31 
\enddata
\tablecomments{Columns list telescope data used for fitting the
  surface brightness profile, subclusters names, best fit for the core radius,
  $\beta$ (Equation (\ref{BS_beta})), and reduced $\chi$-square.}
\label{tab:comp_beta_model}
\end{deluxetable*}


\section{X-ray Observations and data reduction}

\subsection{{\em Chandra}}

We observed G345 on November 21 and 26, 2012, with the \cxc ~X-ray Observatory
(ACIS-I detectors, VF mode, ObsIds 
\dataset[ADS/Sa.CXO#obs/15133]{15133} -- 15 ks -- and
\dataset[ADS/Sa.CXO#obs/15583]{15583} -- 15 ks --
see center panel of Figure \ref{fig:cxc_xmm_rosat}). 
The data were reduced following the processing described in
\citet{2005Vik}, applying the calibration files CALDB 4.5.3. 
The data processing includes corrections for the time dependence of the charge
transfer inefficiency and gain, and a check for periods
of high background (none were found). Also, readout artifacts were
subtracted and standard blank sky background files were used for 
background subtraction. 

\subsection{XMM-{\em Newton}}

We observed G345 on April 2, 2013, with the \xmm ~
Observatory for 22.7 ks (ObsId 0692930101). We used the {\it images}
script\footnote{\scriptsize \href{http://xmm.esac.esa.int/external/xmm_science/gallery/utils/images.shtml}{xmm.esac.esa.int/external/xmm\_science/gallery/utils/images.shtml}}
from the XMM-{\em Newton}
website to create exposure corrected images of the cluster in the
0.5--2.0 keV energy band. This script
removes periods of high
background, as well as bad pixels and columns, spatially smooths and
exposure corrects the data and merges PN and MOS observations (see
Figure \ref{fig:xmm_esored} left panel and Figure \ref{fig:cxc_xmm_rosat}
right panel). 

\newpage

\subsection{ROSAT}

\begin{figure*}[hbt!]
\centerline{%
\includegraphics[width=1.05\textwidth]{%
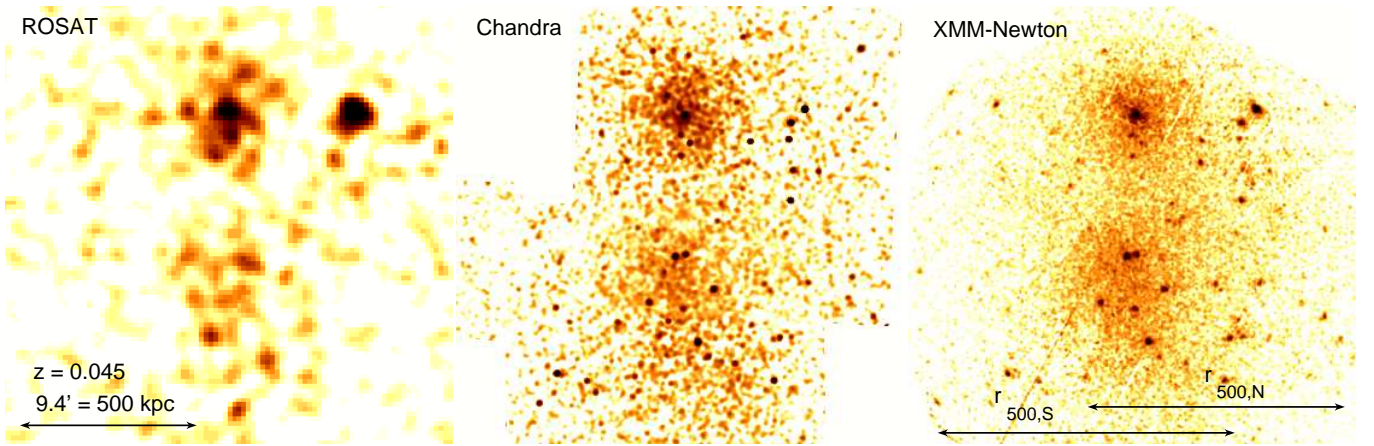}}
\caption{
\small{
Left: The 0.4--2.4 keV ROSAT PSPC image of
G345. Center: The 0.5--2.0 keV \cxc ~image of G345. Right:  The 0.5--2.0 keV
XMM-{\em Newton} image of G345 showing $r_{500}$ for each subcluster. All three images have the same  spatial scale.}
}
\label{fig:cxc_xmm_rosat}
\end{figure*}

G345 was observed on April 2, 1991 for 2.7 ks (ObsId RP800042A00) and on April 1, 1992
for 1.6 ks (ObsId RP800042A01) with the ROSAT PSPC. We used available hard band (0.4--2.4 keV) images,
background, and exposure maps to create a background subtracted,
exposure corrected image of the cluster that we used to
constrain the subclusters' surface brightness profiles at large
radii (see Figure \ref{fig:cxc_xmm_rosat} left panel). 

\section{Spatial and Spectral Analysis}


\subsection{X-ray Surface Brightness Radial Profiles}\label{sec.SB}

\begin{figure*}[hbt!]
\centerline{
\includegraphics[width=0.5\textwidth]{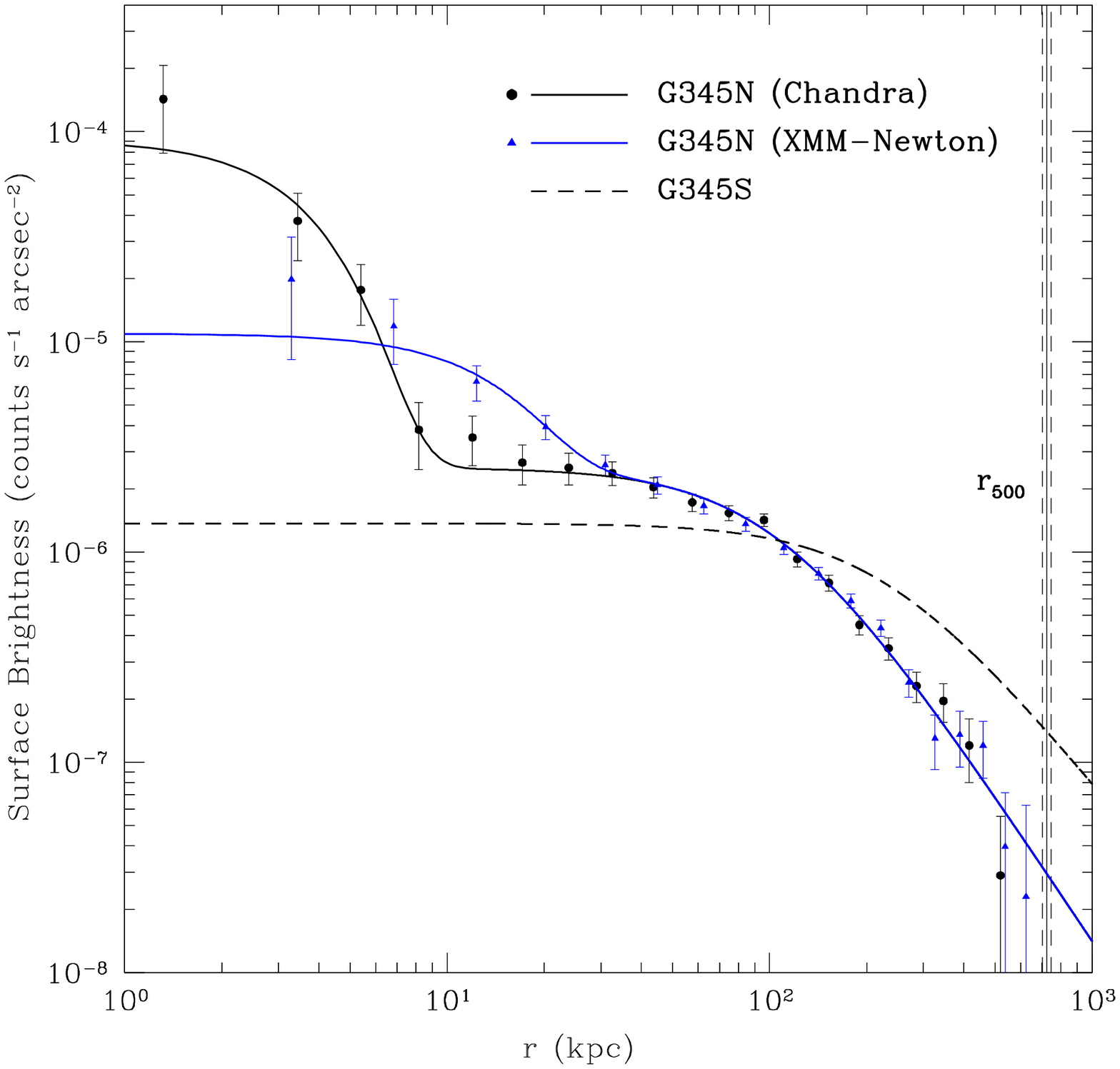}
\includegraphics[width=0.5\textwidth]{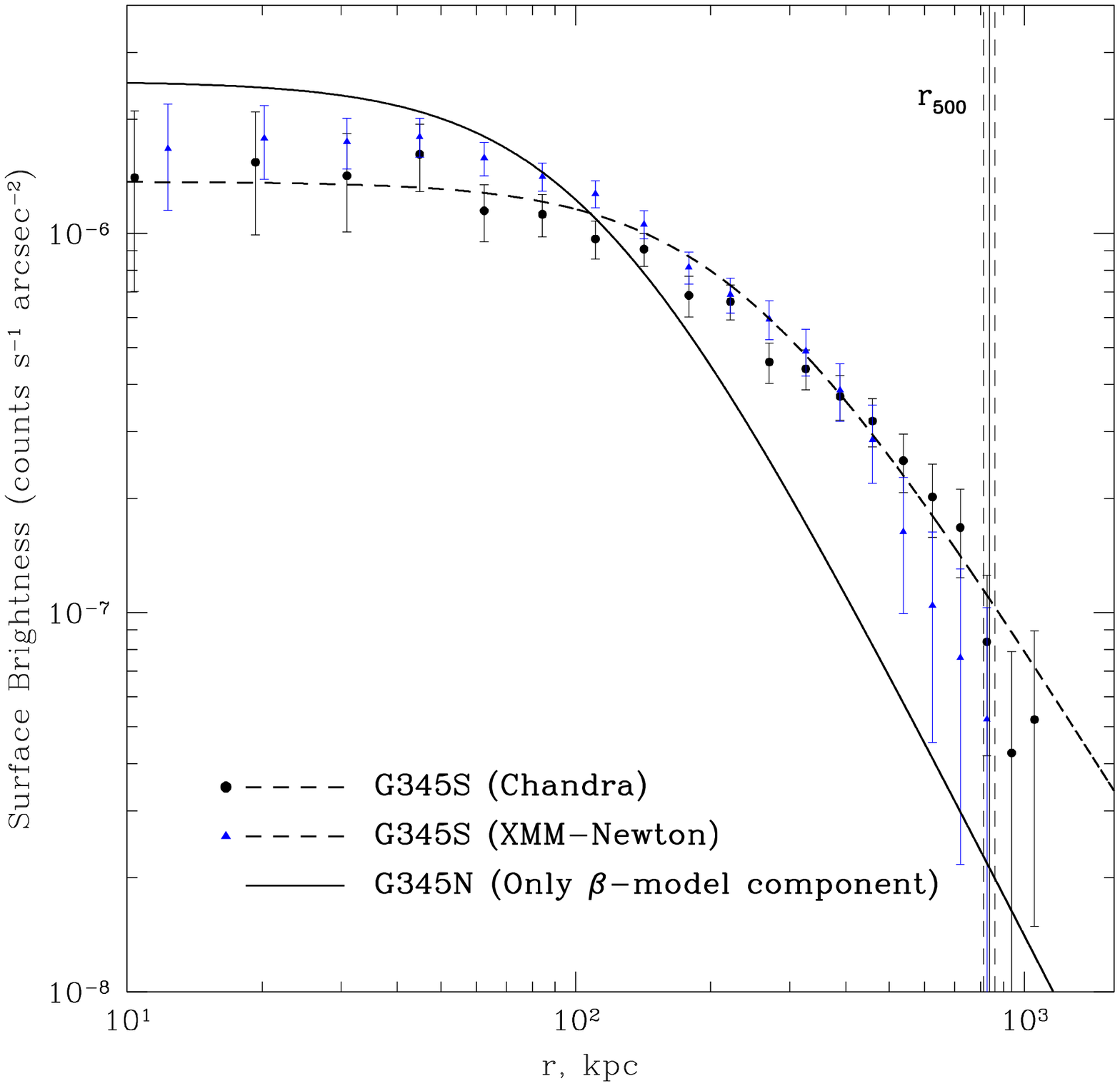}}
\caption{
\small{
Left: X-ray background-subtracted surface brightness profiles and fits for G345N (solid lines), along
with the surface brightness fit for G345S (dashed
line). The difference between the profiles in the core is due to the 
XMM-{\em Newton} poorer angular resolution compared to that of
  {\em Chandra}. The Gaussian function in
the \cxc ~fit describes the cool galactic corona. For
XMM-{\em Newton} the Gaussian function to describe the core is broader, corresponding to the telescope's larger PSF. Right: X-ray surface
brightness profile of G345S (dashed line), along
with the surface brightness fit for the G345N cluster component (solid line)
for comparison.
For each profile, regions in the direction of the other subcluster are
excluded. For display purpose, the XMM-{\em Newton} central surface brightness
  in the $\beta$-model was normalized to match the central surface
  brightness in the {\em Chandra} fit.  The background level is $(1.70
  \pm 0.13) \times 10^{-7} \rm ~counts~s^{-1}~arcsec^{-2}$  (normalized
  to the {\em Chandra} ACIS-I count rate sensitivity in the 0.5--2.0
  keV energy band).
}
}
\label{fig:sb_plot}
\end{figure*}

The surface brightness is the projection of the plasma emissivity
along the line of sight. 
We fit the X-ray surface brightness radial profile of each G345 subcluster with a
$\beta$-model \citep{1976Cavaliere} which is well suited for
non-cool-core relaxed clusters. This model is defined as:
\begin{equation}
S(r)= S_{0} \left[ 1 + 
       \left(\frac{r}{r_{c}}\right)^{2} \right]^{-3\beta+0.5},
\label{BS_beta}
\end{equation}
where $r_{c}$ is the core radius, $\beta$ is the shape parameter, and
$S_0$ is the central surface brightness. 
Figure \ref{fig:sb_plot} shows the surface brightness profiles for the northern (left panel)
and southern (right panel) subclusters, along with the best
$\beta$-model fits.
We detect X-ray
emission from {\em Chandra} and XMM-{\em Newton} to a radius of $\sim$ 700 kpc for the northern component
and to $\sim$ 1 Mpc for the southern one. The background level is $(1.70
  \pm 0.13) \times 10^{-7} \rm ~counts~s^{-1}~arcsec^{-2}$ (normalized
  to the {\em Chandra} ACIS-I count rate sensitivity in the 0.5--2.0
  keV energy band).
For G345N, we modeled the X-ray surface brightness with the
addition of a Gaussian function to describe the X-ray emission associated
with the central giant elliptical (ESO 187-G026). The Gaussian function in
the \cxc ~fit describes the bright core, whereas in
the case of XMM-{\em Newton} fit, the Gaussian function is broader due to the
  telescope's larger PSF compared to {\em Chandra}.
With {\em Chandra}, we extracted a spectrum from a circle of 7 kpc (7.9$''$)
in radius centered on ESO 187-G026 to estimate the gas temperature of the bright
central region  (using the spectrum from
an annulus with inner and outer radii of 7 (7.9$''$) and 27 kpc
(30.5$''$) respectively, as the background component to properly subtract the
cluster emission). For the galaxy, we obtain $kT = 0.92^{+0.13}_{-0.15}$ keV and
$L_{\rm X} = 2.62^{+0.02}_{-1.16} \times 10^{41} \rm ~ erg~s^{-1}$. In
the Appendix we show that the 
(stellar + LMXB) luminosity of this galaxy is 
$L_{\rm 0.5-2.0~keV} = 1.74 \times 10^{40} \rm ~erg~s^{-1}$
which is only $\sim 7\%$ of the total 0.5--2.0 keV
luminosity within 7 kpc (7.9$''$)
of its center. Thus, the majority of the X-ray emission
comes from diffuse gas.
In a similar way, \citet{2001Vik} showed that there are $\sim$ 3 kpc cores of
1--2 keV gas in both giant ellipticals (NGC4874 and NGC4889) 
in the massive and hot ($kT \sim 8$ keV) Coma cluster, and 
\citet{2007Sun} presented a systematic investigation of X-ray
thermal coronae in a survey of 25 nearby clusters. They showed that
cool galactic coronae ($kT$ = 0.5--1.0 keV generally) are common in cluster cores. 
Using ROSAT's larger field of view to better constrain
the background level along with {\em Chandra} and XMM-{\em Newton} observations, we fitted each subcluster X-ray surface brightness to a 
$\beta$-model profile and determined $\beta$ and $r_{\rm c}$ (the background level for each surface brightness
profile ({\em Chandra}, XMM-{\em Newton}, and ROSAT)  was free to vary independently, while
$r_{\rm c}$ and $\beta$ were tied to a single best fit value). For G345N, we obtained $\beta$ = \betan ~ and
$r_{\rm c}$ = \rcn ~ kpc {\bf ($\chi^2_{\rm red} =
  1.07$)}, 
and for G345S, we obtained 
$\beta$ = \betas ~ and $r_{\rm c}$ = \rcs ~ kpc  {\bf ($\chi^2_{\rm red} =
  1.31$)}. 
We also fit the {\em Chandra} and XMM-{\em Newton} profiles individually and found consistent 
$\beta$ and $r_{\rm c}$.
The best fit parameters are summarized in Table \ref{tab:beta_model}. 
The southern subcluster has a flatter $\beta$ and a larger core radius than the northern 
subcluster, which along with the poorer $\beta$-model fit and higher central entropy of G345S 
(the entropy profiles for each subcluster will be discussed in Section \ref{sec:entropy_profile}) 
suggest this subcluster is still in the process of forming.
However, visual inspection of Figure \ref{fig:sb_plot} as well as the best
fit reduced  $\chi^2$ show that the
$\beta$-model
describes well the X-ray surface brightness profiles of G345N (with the exception
of the central region due to the central galaxy's cool galactic
corona X-ray emission) and G345S,
validating the assumption that the gas density defined by the surface
brightness follows this model. 

\subsection{Gas Density Radial Profiles}

The $\beta$-model gas density distribution that corresponds to the
surface brightness distribution given by Equation (\ref{BS_beta}) is:
\begin{equation}
\rho_{\rm g} (r) = \rho_{\rm g,0}
\left[1+\left(\frac{r}{r_{\rm c}}\right)^2\right]^{-3\beta /2}, 
\label{density}
\end{equation}
where $\rho_{\rm g,0}$ is the central gas density. The core radius, $r_{\rm c}$, and
$\beta$ are constrained from fitting the X-ray surface
brightness profile.

The gas mass within $r_{500}$ (the radius defining a sphere
whose interior mean mass density is 500 times the critical density at the
cluster redshift -- see Section \ref{sec:total_mass}) is then given by:
\begin{eqnarray}
M_{\rm g,500} && = 4 \pi \rho_{g,0}  \int_{0}^{r_{500}} 
\left[1+\left(\frac{r}{r_{\rm c}}\right)^2\right]^{-3\beta /2} r^2 dr
\nonumber \\
&& = \frac{4\pi}{3} \rho_{g,0} r_{500}^3~
_2F_1(3/2,3\beta/2;5/2;-y^2), 
\label{gas_mass}
\end{eqnarray}
where $y=r_{500}/r_{\rm c}$ and $_2F_1(a,b;c;d)$ is the Gauss
hypergeometric function. 
For a plasma with a given electron to hydrogen number densities ratio 
$n_{\rm  e}/n_{\rm H}$,  the central electron number density, $n_{\rm  e,0}$ 
is calculated as:
\begin{eqnarray}
&&n_{\rm e,0}  = \left(
\frac{K (n_{\rm e}/n_{\rm H}) [D_{\rm A}(1+z)]^2  (6\beta-3) \times 10^{14}}
{2^{6\beta-3} r_{\rm c}^3 B(3\beta-1/2,3\beta-1/2)}
\right)^{1/2} \times \nonumber \\ 
&&\left(\left[1+\left(\frac{R_{\rm i}}{r_{c}}\right)^{2}
\right]^{-3\beta+3/2}-\left[1+\left(\frac{R_{\rm f}}{r_{c}}\right)^{2} \right]^{-3\beta+3/2}\right)^{-1/2}
\label{n0}
\end{eqnarray}
where $D_{\rm A}$ is the
angular distance of the cluster, $B(a,b)$ $a,b \in$ $\mathbb{R}$  is the 
Euler $\beta$ function, $R_{\rm i}$ and $R_{\rm f}$ are the projected
initial and final radii of the annulus for which the spectrum has been
fit and the normalization, $K$, was computed using the \textsc{apec}
model in XSPEC. The relation between the gas density and electron
number density is given by $\rho_{\rm g} =
\mu_{\rm e} n_{\rm e} m_{\rm a}$, where $\mu_{\rm e}$ is the mean
molecular weight per electron and $m_{\rm a}$ is the atomic mass unit. For a metallicity of
0.3 $Z_\odot$, using the reference values from \citet{1989AndersGrevesse} we obtain 
$\mu_{\rm e} = 1.1706$ and $n_{\rm e}/n_{\rm H} = 1.1995$.
The central electron number and gas densities for G345N are
approximately four times higher than those for G345S. These are given in
Table \ref{tab:beta_model}.


\subsection{Gas Temperature Radial Profiles}

The temperature profiles of most clusters have
a broad peak within 0.1--0.2 $r_{200}$ and decreases at larger radii,
reaching approximately
50\% of the peak value near $0.5\,r_{200}$ \citep{2006Vik}. In cool-core clusters there is also a temperature
decline toward the cluster center due to
radiative cooling. The analytic model constructed by \citet{2006Vik}
for the 3D temperature profile describes these general features.
However, due to the quality of our data, we employed a simplified form
of this temperature profile with some of the parameters fixed
(the universal temperature profile from \citet{2006Vik}, but with
the transition radius, $r_t$, allowed to vary). Thus

%
\begin{equation}\label{eq:tprof_mod}
  T_{\rm 3D}(r) =  \frac{T_0}{[1+(r/r_t)^2]} \times 
  \frac{[r/(0.075 r_t)]^{1.9}+0.45}{[r/(0.075 r_t)]^{1.9}+1}. 
\end{equation}

Monte-Carlo simulations were performed to estimate the uncertainties
in the best fit values for the parameters in this analytical model.
This analytic model for $T_{\rm 3D}(r)$ (Equation (\ref{eq:tprof_mod})), allows
very steep temperature gradients. In some realizations,
such profiles are formally consistent with the observed projected
temperatures because projection flattens steep gradients.  However,
steep values of $dT/dr$ often lead to unphysical mass estimates, for example, 
profiles with negative dark matter density at some radii. We addressed this problem in the
Monte-Carlo simulations by accepting only those realizations in which the
best-fit $T(r)$ leads to $\rho_{\rm tot}>\rho_{\rm gas}$ in the radial
range $r \leq 1.5\,r_{500}$.
Finally, in the same radial range, we
verified that the temperature profiles corresponding to the mass uncertainty
interval are all convectively stable\footnote{Assuming the motion of a
gas element in a medium in hydrostatic equilibrium is adiabatic, 
one can apply Newton's second law to the net force per unit volume of
the gas to obtain two solutions: an
oscillatory one, which is stable, and a run-away motion, which is
unstable. The stable solution  imposes $d\ln  T/d\ln  \rho_g <
\gamma-1$, where $\gamma$ is the adiabatic index. This is also known as
the Schwarzchild criterion. For $\gamma = 5/3$, $d\ln  T/d\ln  \rho_g < 2/3$.}, i.e., $d\ln  T/d\ln  \rho_g < 2/3$.

\begin{figure*}[hbt!]
\centerline{%
\includegraphics[width=0.33\textwidth]{%
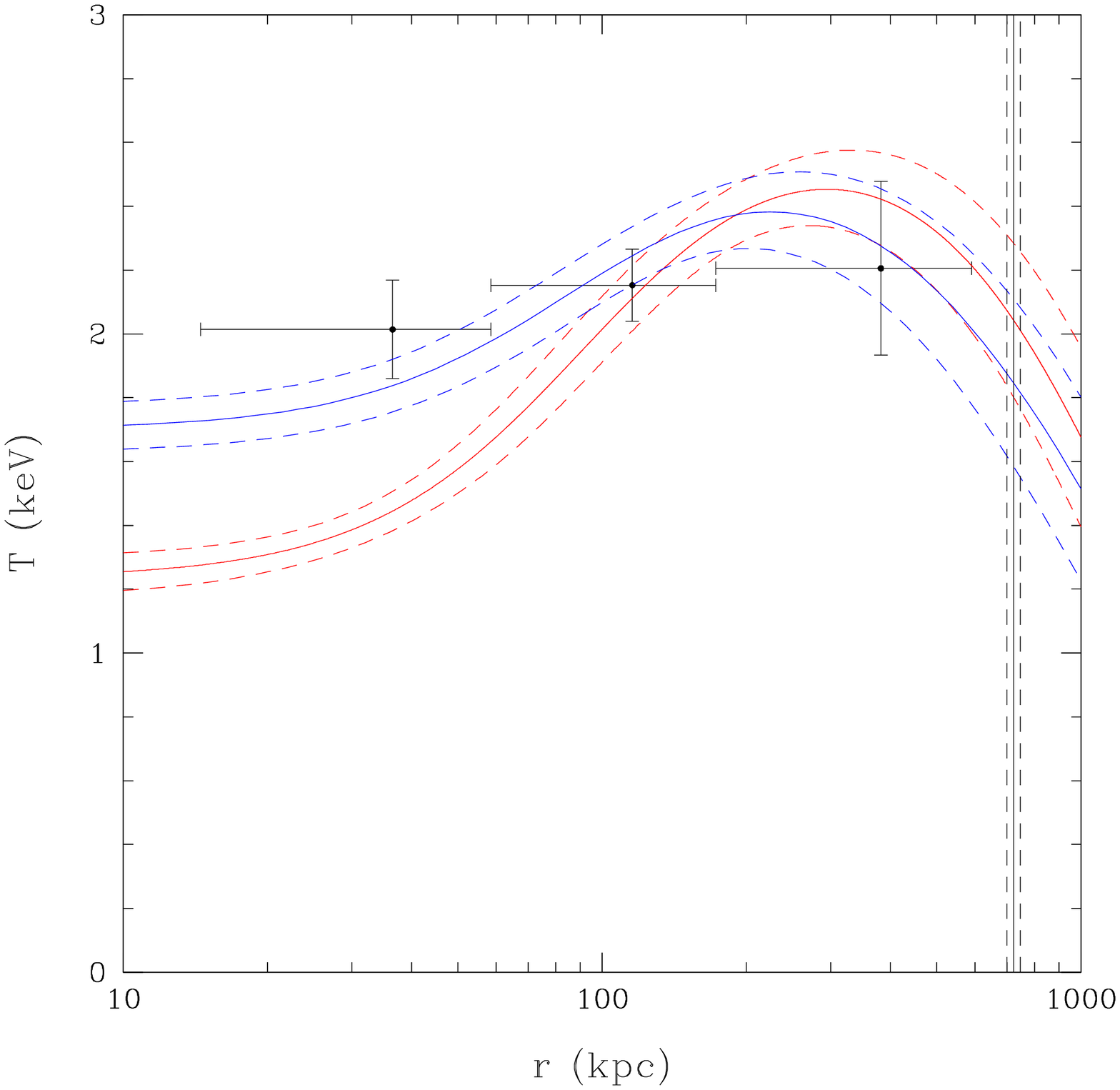}
\includegraphics[width=0.33\textwidth]{%
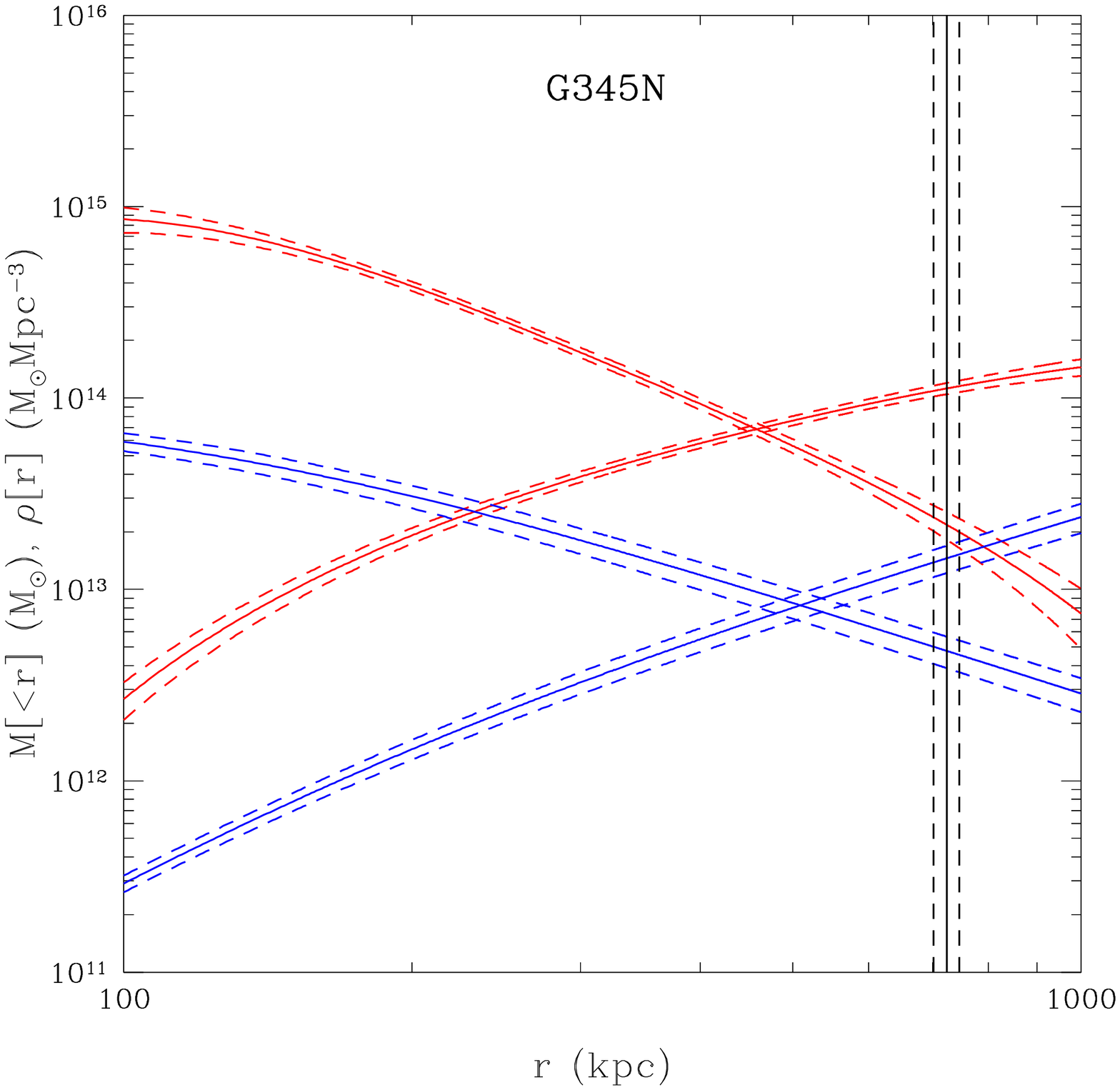}
\includegraphics[width=0.33\textwidth]{%
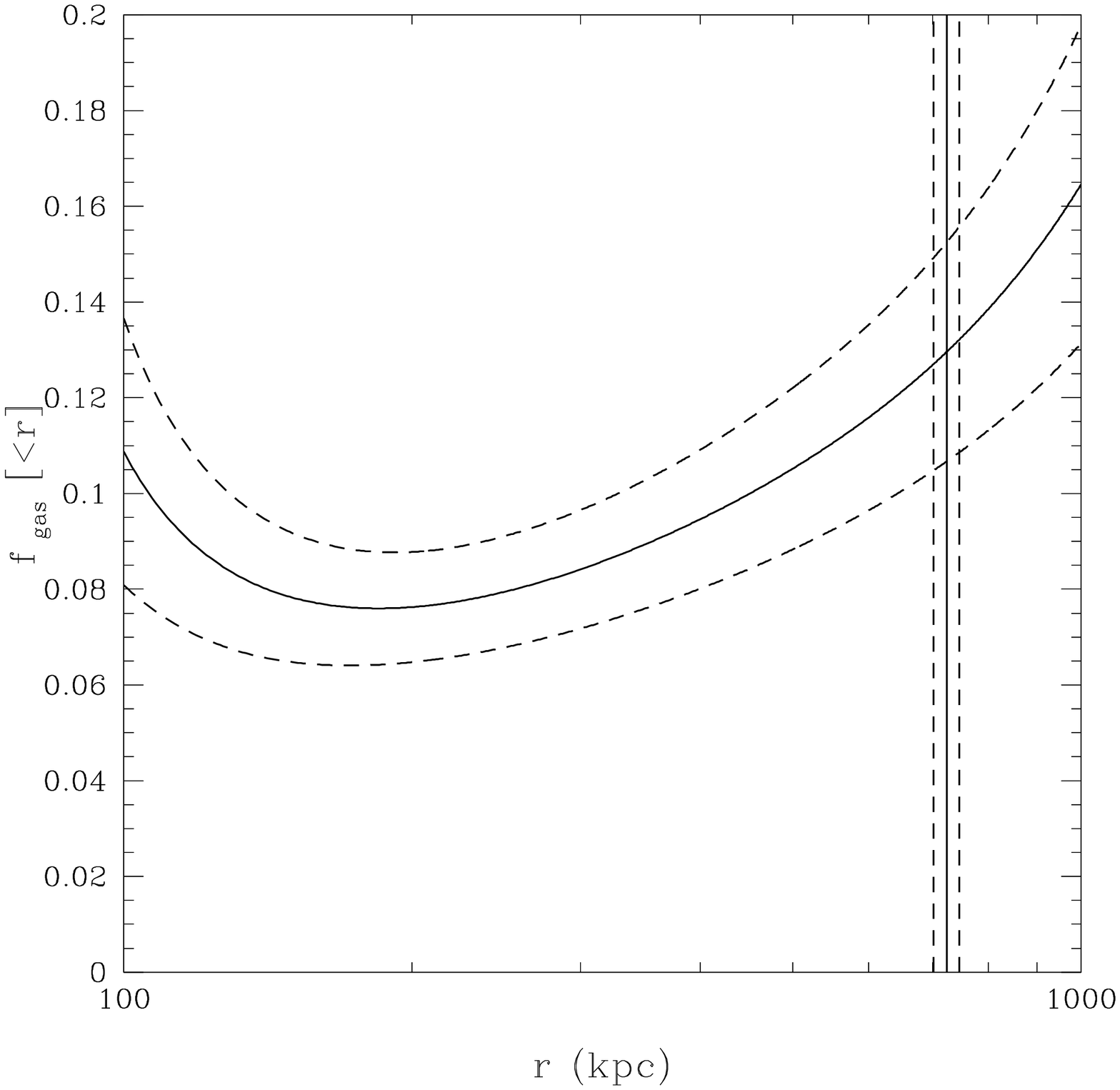}}
\centerline{%
\includegraphics[width=0.33\textwidth]{%
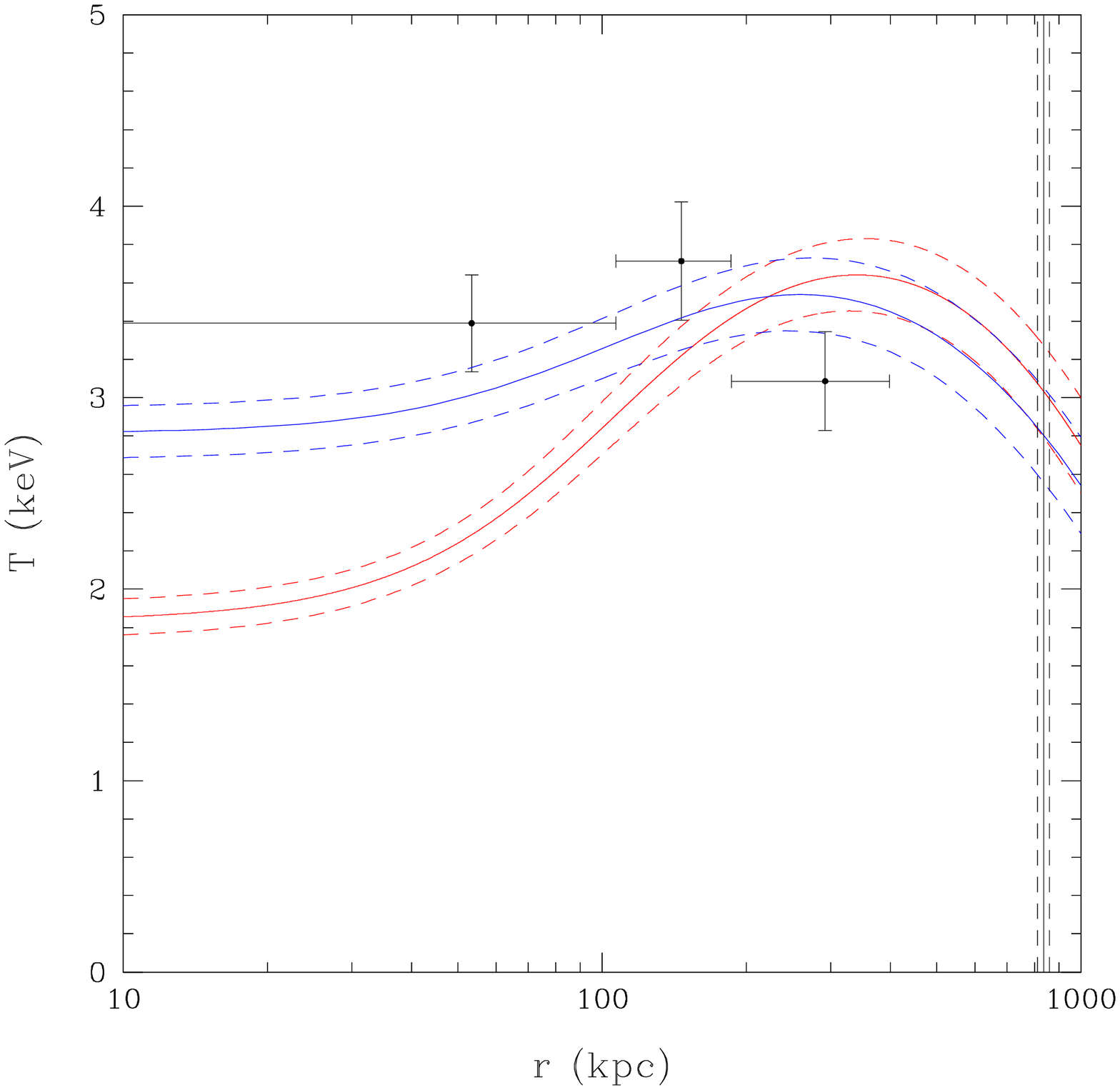}
\includegraphics[width=0.33\textwidth]{%
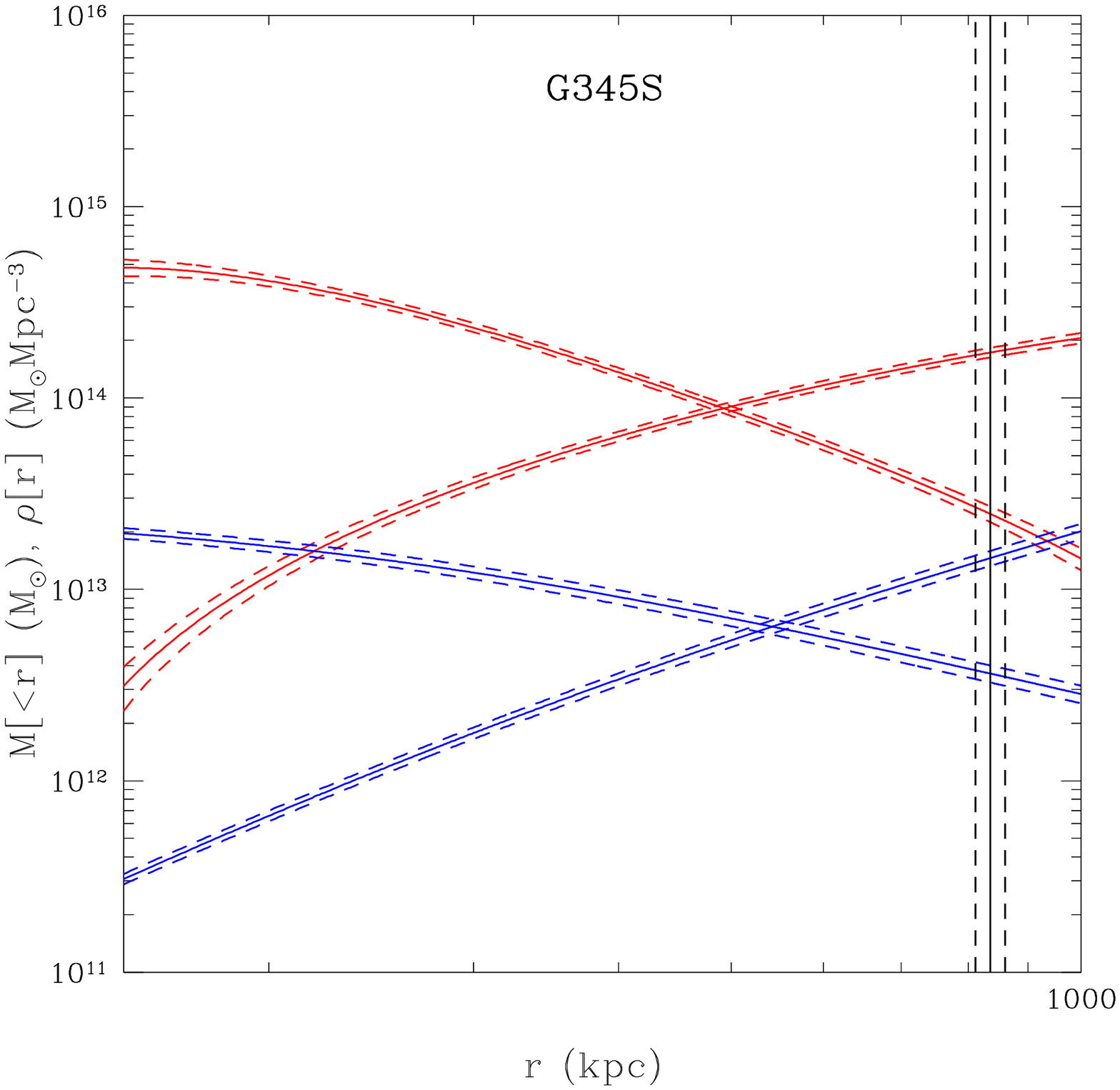}
\includegraphics[width=0.33\textwidth]{%
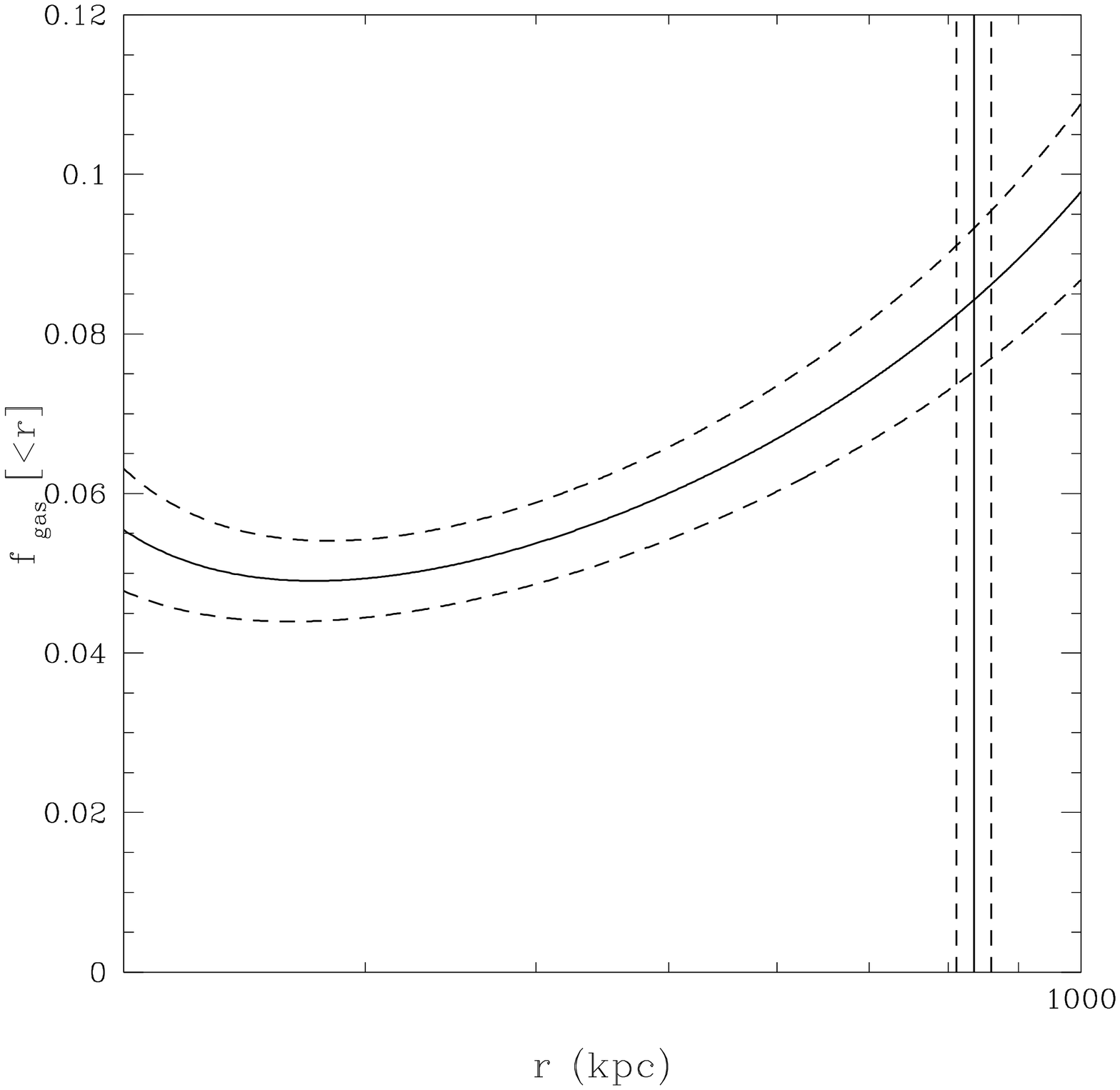}}
\caption{Top: G345N. Bottom: G345S.
Left: Temperature profiles.  Measured
  projected temperatures are shown by circles with error bars. Solid red and blue lines
  show the best fit 3D model and the corresponding projected profile,
  respectively. 
Center: Mass and density profiles ($M(<r)$ increases with radius
  while $\rho(r)$ decreases). Red and blue lines show the results for the total
  mass and gas mass, respectively. 
Right: Gas mass fraction as a
  function of radius. 
Solid line shows the enclosed
$f_{\rm gas}=M_{\rm g}(<r)/M_{\rm tot}(<r)$. 
The vertical line shows the radius $r_{500}$ derived from the
  best fit mass model assuming hydrostatic equilibrium. The dashed
  lines show the 68$\%$ confidence level limits for the best fits. 
}
\label{fig:temp_mass_fg}
\end{figure*}

Using XMM-{\em Newton} observations, we extracted spectra in three
half annuli subtending 180
degrees on the side opposite to the companion subcluster, centered on
each subcluster in the radial range from from 0 to
$\sim$ 400--600 kpc for each subcluster. Each half annulus has at least 3000 source counts. 
We fit these with an absorbed \textsc{apec} model. The measured Galactic hydrogen column
density in the direction of the cluster is $N_{\rm H} = 2.51 \times
10^{20} {\rm ~cm^{-2}}$, which was kept fixed while fitting the spectra.
We then followed the procedures described below to obtain the 2D and
3D temperature profiles. 
The measured 2D, fitted 2D, and 3D temperature profiles are presented
in the left panel of Figure \ref{fig:temp_mass_fg}. The 2D temperature
profile was computed by projecting the 3D temperature weighted by gas density 
using the spectroscopic-like temperature \citep[a formula for the
temperature which matches the spectroscopically
measured temperature within a few percent,][]{2004Mazzotta}:
\begin{equation}
T_{\rm 2D}=T_{\rm spec} \equiv \frac{\int \rho_{\rm g}^2 T_{\rm
    3D}^{1/4} dz}{\int \rho_{\rm g}^2 T_{\rm 3D}^{-3/4} dz}
\label{eq:tspec}
\end{equation}

The spectroscopic-like temperature
was also computed in the (0.15 -- 1)  $r_{500}$ range using
the model for the 3D temperature and the gas density profile.
For G345N, we found $kT = ~ $\kThydn$ $ keV, and for G345S, $kT = ~ $\kThyds $ $ keV.

\begin{deluxetable*}{cccccccc}
\tablecaption{Temperature Determination} 
\tablewidth{0pt} 
\tablehead{ 
\colhead{Subcluster} &
\colhead{$r$ (kpc)} &  
\colhead{$kT$ (keV)} &
\colhead{Subcluster} &
\colhead{$r$ (kpc)} & 
\colhead{$kT$ (keV)} 
}
\startdata 
      & 14.5 -- 58.5 & $2.01 \pm 0.15$ &       &  0 -- 106.8 & $3.39 \pm 0.25$ \\
G345N & 58.5 -- 172.7 & $2.15 \pm 0.11$ & G345S & 106.8 -- 186.0 & $3.71 \pm 0.31$ \\
      & 172.7 -- 591.5 & $2.21 \pm 0.27$ &       & 186.0 -- 398.4 & $3.08 \pm 0.26$
\enddata
\tablecomments{Columns give subclusters' names, radial range for the
  annuli used for temperature extraction, and the measured temperatures with 1$\sigma$ uncertainties.} 
\label{tab:temp_prof}
\end{deluxetable*}

\begin{deluxetable}{ccc}
\tablecaption{Parameters for the Temperature Profile} 
\tablewidth{0pt} 
\tablehead{ 
\colhead{Subcluster} &
\colhead{$T_0$} &
\colhead{$r_{\rm t}$} \\
\colhead{} &
\colhead{(keV)} &
\colhead{(kpc)}
}
\startdata 
G345N & \Tn & \rtn \\
G345S & \Ts & \rts  
\enddata
\tablecomments{Columns list best fit values for $T_0$ and $r_{\rm t}$
given by Equation (\ref{eq:tprof_mod}).} 
\label{tab:temp_prof}
\end{deluxetable}


\section{Total and Gas Masses of Subclusters}\label{sec:total_mass}

Given the three-dimensional models for the gas density and temperature
profiles, the total mass within a radius $r$ can be computed
for each subcluster from the equation for hydrostatic equilibrium (e.g., Sarazin
1988),
\begin{eqnarray}\label{e_m_r}
M(r) &=& \frac{-kTr}{\mu m_H G} 
\left( 
\frac{d\ln \rho_{\rm g}}{d\ln r}+
\frac{d\ln T}{d\ln r}
\right) \nonumber \\ 
&=& -3.67 \times 10^{13} ~ M_\odot kT r 
\left( 
\frac{d\ln \rho_{\rm g}}{d\ln r}+
\frac{d\ln T}{d\ln r}
\right), 
\end{eqnarray}
where $k$ is the Boltzmann constant, $T$ is the gas temperature in units of K, and $r$ is in units of Mpc (the normalization
corresponds to  $\mu = 0.6107$, appropriate for 0.3 $Z_\odot$).  

Using Equation (\ref{e_m_r}), $r_{500}$  is computed by
solving 
\begin{eqnarray}
M(r_{500})=500 \rho_c (4\pi/3) r_{500}^{3},\label{m500_def}
\end{eqnarray}
where 
$\rho_c$ is the critical density of the Universe at the cluster
redshift. For G345N, we
obtain $r_{500} = ~ $\rhydn$ $ kpc 
and a corresponding hydrostatic mass of 
$M_{\rm 500,hyd} = ~ $\mhydn$ M_\odot$. For G345S,
we find $r_{500} =  ~ $\rhyds$ $ kpc with a corresponding hydrostatic mass of 
$M_{\rm500,hyd} =  ~ $\mhyds$ M_\odot$.
Using the best fit parameters for the $\beta$-model (see Table
\ref{tab:beta_model}), we compute a gas mass within $r_{500}$ of $M_{\rm g,500} = ~ $\ghydn$
M_\odot$ for G345N, and $M_{\rm g,500} = ~ $\ghyds$  M_\odot$ for G345S.

The gas and total masses and density profiles, as well as the gas mass fractions within
$r_{500}$ for both G345N and
G345S are shown in the center and right panels of Figure \ref{fig:temp_mass_fg}. The gas mass
fraction within $r_{500}$ is 
$f_{\rm g} = ~ $\fghyds$ $ for G345S and is in agreement with the expected value
from \citet{2006Vik} for clusters $\sim$ 1--2 $\times 10^{14} M_\odot$
($f_{\rm g} \sim 0.09$), whereas the gas mass fraction for G345N
is slightly larger, $f_{\rm g} = $\fghydn$ $. We measure the X-ray luminosity
within 245 kpc and extrapolate to $r_{500}$ for both subclusters,
using the best fit $\beta$-model parameters. For G345N, we obtain a
bolometric X-ray luminosity $L_{X,r_{500}} = (2.65 \pm 0.37) \times 10^{43} \rm ~erg~s^{-1}$, and
for G345S, we obtain $L_{X,r_{500}} = (3.72 \pm 0.51) \times 10^{43} \rm ~erg~s^{-1}$.
These results are summarized in Table \ref{tab:hyd_eq}. 

Alternatively, if we use the gas mass and temperature, 
the total mass can be estimated from the $Y_X$--$M$ scaling relation of \citet{2009Vik},
\begin{eqnarray}
M_{\rm 500,Y_X} = E(z)^{-2/5}A_{\rm YM}\left(\frac{Y_{\rm X}}{3\times10^{14}M_\odot
  {\rm keV}}\right)^{B_{\rm YM}}, \nonumber \\
\label{e_yx_m}
\end{eqnarray}
where $Y_{\rm X} = M_{\rm g,500} \times kT_{\rm X}$, $M_{\rm
  g,500}$ is given by
Equation (\ref{gas_mass}), and $T_{\rm X}$ is the spectroscopic-like
temperature (see Equation (\ref{eq:tspec})) in the (0.15--1) $\times ~ r_{500}$ range,
computed based on the model for the gas density and 3D temperature profiles. 
$A_{\rm YM}=5.77\times10^{14}h^{1/2}  M_\odot$ and $B_{\rm YM}=0.57$ \citep{2012Maughan}. Here,
$M_{\rm Y_X,500}$ is the total mass within $r_{500}$,
and $E(z)=[\Omega_{\rm M}(1+z)^3 + (1-\Omega_{\rm M}-\Omega_\Lambda)(1+z)^2 +
\Omega_\Lambda]^{1/2}$ is the function describing the redshift evolution of the Hubble
parameter. As for the mass determination assuming hydrostatic equilibrium, we estimated 1$\sigma$ uncertainties in the $Y_{\rm X}$ derived quantities 
using Monte Carlo simulations. We also
added to the Monte Carlo procedure a 1$\sigma$ systematic uncertainty of 9\% in
the mass determination, as discussed by \citet{2009Vik}.

Using Equations (\ref{m500_def}) and (\ref{e_yx_m}), we compute
$r_{\rm 500,Y_x}$ = \ryxn ~kpc for G345N
with a corresponding total mass of 
$M_{\rm 500,Y_x} = $\myxn$ M_\odot$ and gas mass
$M_{\rm g,500,Y_X} = ~ $\gyxn$ M_\odot$.
For G345S, we obtain $r_{\rm 500,Y_x}$ = \ryxs ~ kpc
with a corresponding total mass of 
$M_{\rm 500,Y_x} = ~ $\myxs$ M_\odot$, and gas mass 
$M_{\rm g,500,Y_X} = ~ $\gyxs$ M_\odot$, in agreement with
results obtained assuming the cluster to be in hydrostatic equilibrium.

Using the $Y_{\rm X}$ relation, the gas mass fractions within $r_{500}$
are $f_{\rm g} = ~ $\fgyxn$ $ for G345N and $f_{\rm g} =  ~ $\fgyxs$ $ for
G345S. These values are consistent with gas fractions computed above
assuming hydrostatic equilibrium. These results are summarized in Table \ref{tab:yx}.

\begin{deluxetable*}{cccccccc}
\tablecaption{Physical Properties Derived Assuming Hydrostatic Equilibrium} 
\tablewidth{0pt} 
\tablehead{ 
\colhead{Subcluster} &
\colhead{$r_{\rm 500,hyd}$} &
\colhead{$M_{\rm g,500,hyd}$} &
\colhead{$M_{\rm 500,hyd}$} &
\colhead{$f_{\rm gas,hyd}$} &
\colhead{$kT_{\rm spec}$} &
\colhead{$L_{\rm X,245}$} &
\colhead{$L_{\rm X,r_{500}}$} \\
\colhead{} &
\colhead{(kpc)} &
\colhead{($M_\odot$)} &
\colhead{($M_\odot$)} &
\colhead{} &
\colhead{(keV)} &
\colhead{($10^{43} \rm ~erg~s^{-1}$)} &
\colhead{($10^{43} \rm ~erg~s^{-1}$)}
} 
\startdata 
G345N & \rhydn & \ghydn & \mhydn  & \fghydn & \kThydn & $1.35 \pm 0.08$ & $2.65 \pm 0.37$ \\ 
G345S & \rhyds & \ghyds & \mhyds  & \fghyds & \kThyds & $1.01 \pm 0.06$ & $3.72 \pm 0.51$
\enddata
\tablecomments{Columns list the cluster $r_{500}$, gas mass, total mass derived from
the hydrostatic equilibrium equation (Equation (\ref{e_m_r})), gas
fraction, spectroscopic-like temperature within (0.15 -- 1) $\times ~ r_{\rm 500}$,
and bolometric X-ray luminosities within 245 kpc radii and extrapolated to within $r_{500}$.}
\label{tab:hyd_eq} 
\end{deluxetable*}

\begin{deluxetable*}{cccccccccccc}
\tablecaption{Physical Properties Derived from the $Y_{\rm X}$--$M$
  scaling relation} 
\tablewidth{0pt} 
\tablehead{ 
\colhead{Subcluster} &
\colhead{$r_{\rm 500,Y_X}$} &
\colhead{$M_{\rm g,500,Y_X}$} &
\colhead{$M_{\rm 500,Y_X}$} &
\colhead{$f_{\rm gas,Y_X}$} &
\colhead{$kT_{\rm spec}$} &
\colhead{$L_{\rm X,245}$} &
\colhead{$L_{\rm X,r_{500}}$} \\
\colhead{} &
\colhead{(kpc)} &
\colhead{($M_\odot$)} &
\colhead{($M_\odot$)} &
\colhead{} &
\colhead{(keV)} &
\colhead{($10^{43} \rm ~erg~s^{-1}$)} &
\colhead{($10^{43} \rm ~erg~s^{-1}$)}
} 
\startdata 
G345N & \ryxn & \gyxn & \myxn  & \fgyxn & \kTyxn & $1.35 \pm 0.08$ & $2.72 \pm 0.44$\\ 
G345S & \ryxs & \gyxs & \myxs  & \fgyxs & \kTyxs & $1.01 \pm 0.06$ & $3.58 \pm 0.48$
\enddata
\tablecomments{Columns list the cluster $r_{500}$, gas mass, total mass derived from
the $Y_X$ parameter (Equation (\ref{e_yx_m})), gas fraction,
spectroscopic-like temperature within (0.15 -- 1) $\times ~ r_{\rm 500}$,
and bolometric X-ray luminosities within 245 kpc radii and extrapolated to within $r_{500}$.}
\label{tab:yx} 
\end{deluxetable*}


\section{Subcluster Entropy Profiles}\label{sec:entropy_profile}

The entropy index of the intracluster gas is defined as
\begin{eqnarray}
K = \frac{kT}{n_{\rm e}^{2/3}},
\end{eqnarray} 
where $k$ is the Boltzmann constant, $T$ is the gas temperature, and $n_{\rm e}$ is the
electron density. The entropy profile reflects the
thermodynamic history of the cluster. The entropy increases when
heat energy is deposited into the ICM, and decreases when
radiative cooling carries heat energy away \citep{2005Voit}.  

\begin{figure*}[htb!]
\centerline{\includegraphics[width=0.5\textwidth]{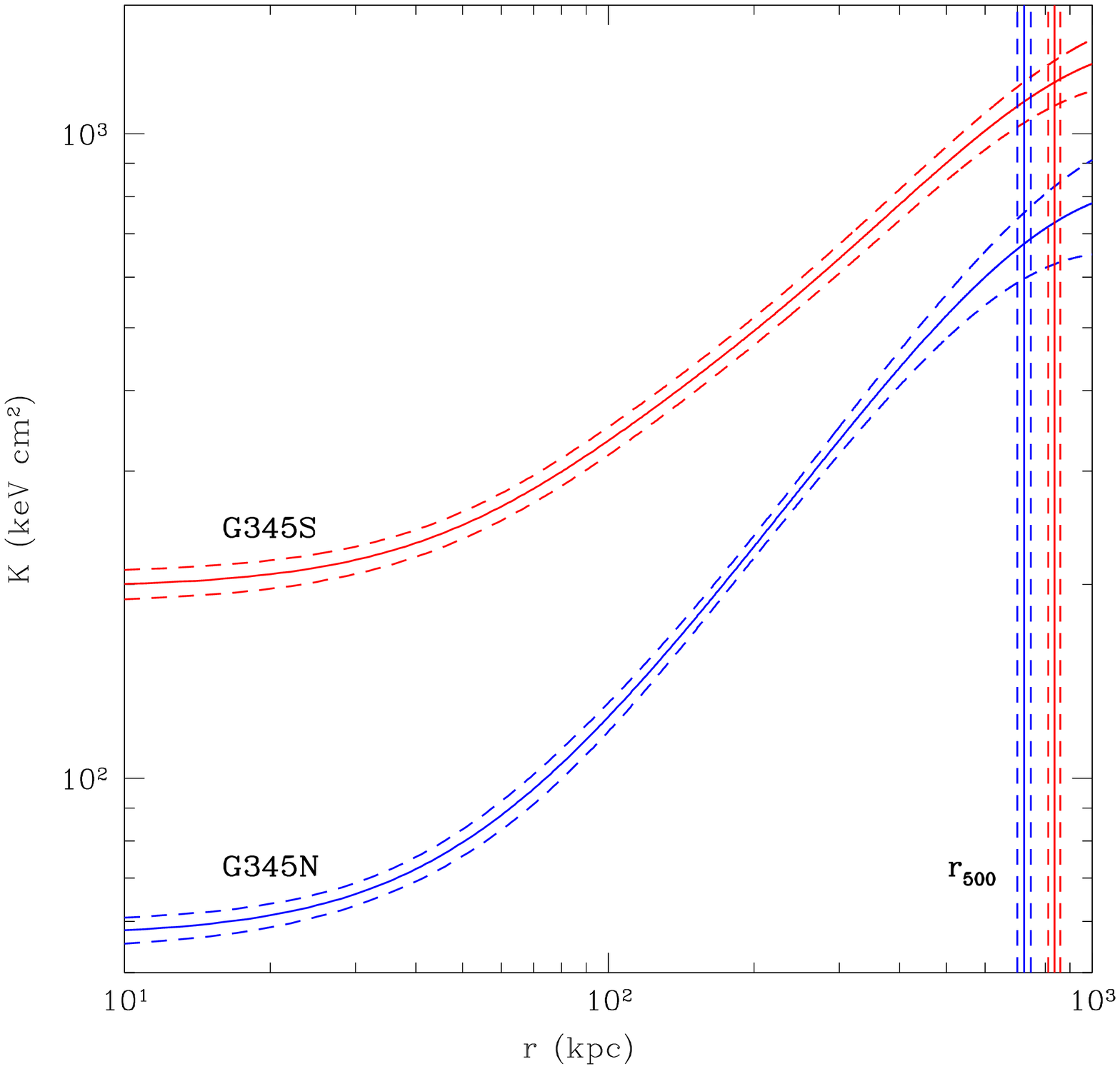}
\includegraphics[width=0.5\textwidth]{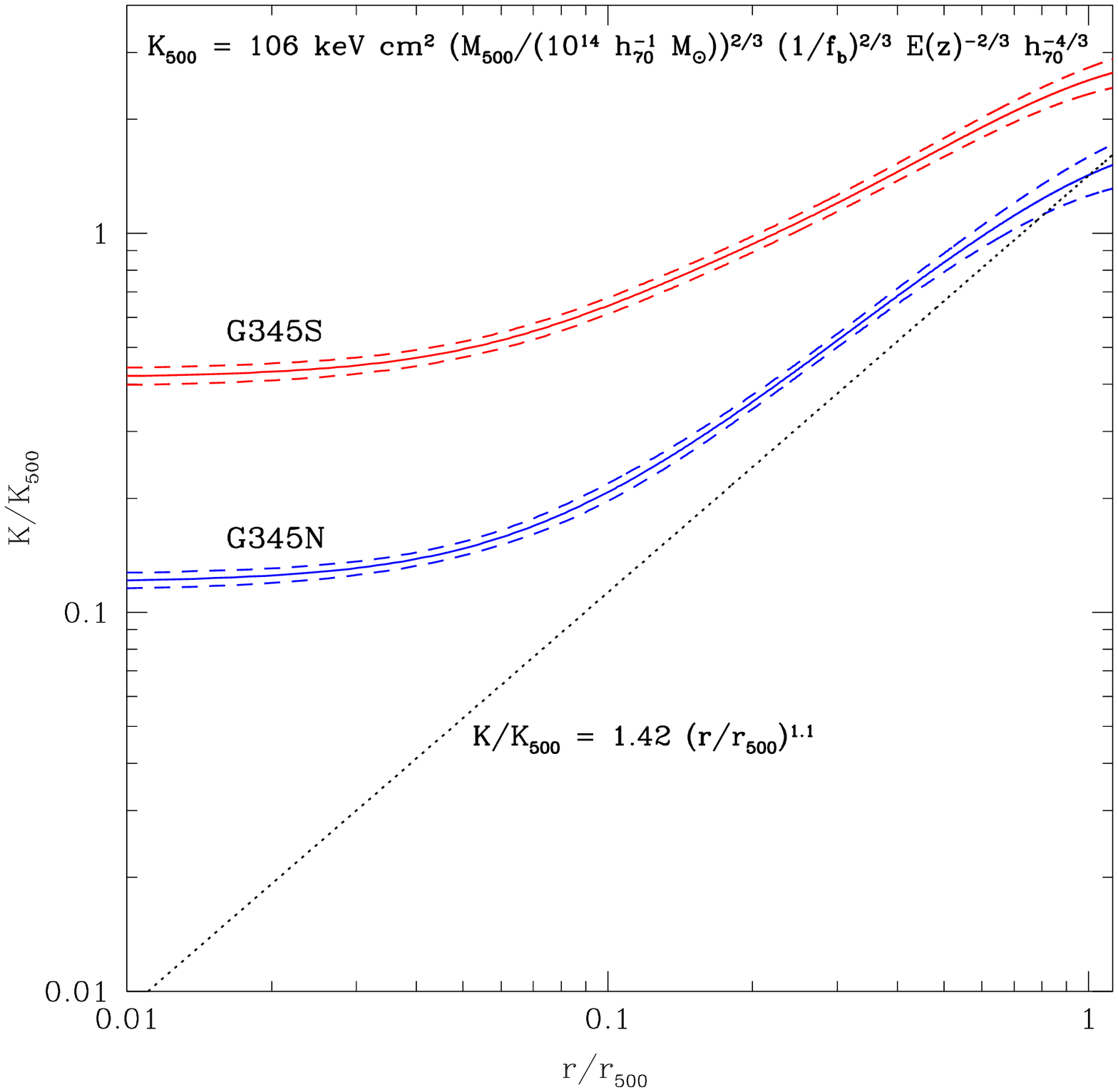}}
\caption{
\small{
Left: Entropy profiles for G345N and G345S. They are remarkably
different. G345S presents a very high central entropy profile. The
corona of the central galaxy has been removed from the profile for
G345N. Right: Dimensionless entropy profiles for G345N and G345S. We see
that the entropy index of G345N is very close to the scaling relation
$K/K_{500} = 1.42 (r/r_{500})^{1.1}$ at $r_{500}$. On the other
hand, G345S exceeds this value at $r_{500}$, suggesting
that non-gravitational processes are playing a significant role in the
thermodynamics of the ICM even at large distances from the center of
the cluster.
}
}
\label{fig:entropy_prof}
\end{figure*}

To better understand the thermodynamic history of the G345 subclusters, we
computed their entropy profiles, which are presented in Figure
\ref{fig:entropy_prof}. They are remarkably
different. G345S has extremely high entropy in its core ($\sim 200 \rm~keV~cm^2$) due to the
extremely low gas density for its observed $\sim$ 3 keV temperature
\citep[typically, the central entropy of a relaxed cluster with $\sim 3$ keV
gas temperature would be $\sim$ 10--30 $\rm ~keV~cm^2$,][]{2005Voit,2013McDonald}.

Here, we present two heating mechanisms that could
explain an entropy increase in the ICM. In the first,
clusters that have experienced recent major mergers
can have high central entropies due to energy deposited in the ICM
through shocks during the merger.
The shock heating will produce an entropy increase, in the case of a
merger of clusters of similar masses, that can be directly related to
the Mach number by:
\begin{eqnarray}
\ln \left(\frac{K_{\rm f}}{K_{\rm i}}\right) &=& 
\ln \left[1+\frac{2\gamma(M^2-1)}{\gamma+1}\right] \nonumber \\ 
&-& \gamma \ln
\left[\frac{(\gamma+1)M^2}{(\gamma+1)+(\gamma-1)(M^2-1)}\right],
\label{eqn:shock_entropy}
\end{eqnarray}
where $K_{\rm i}$ and $K_{\rm f}$ are the initial and final entropies,
respectively \citep{1967Zeldovich}. 

Assuming that the central entropy index of
G345S was similar to that of
G345N before it was perturbed, one can 
put a well-defined requirement on the shock strength in
the core to boost the central entropy by a factor of 3.47 (ratio of
central entropies between the southern and northern subclusters). 
Inserting this entropy ratio into Equation
(\ref{eqn:shock_entropy}),
one obtains a high
Mach number of  $M = 4.8$, which corresponds to $\sim 4300
\rm ~km~s^{-1}$ for a 3 keV cluster. 
This result is implausibly high (assuming that the  mass density
  profile of G345S is described by a NFW profile with
concentration parameter of 4, a point mass falling from infinity would
arrive in the center of G345S with a velocity of $\sim
2600 \rm ~km~s^{-1}$). However, 
shocks are not the only way to raise the entropy in a
merger. Dissipation of turbulence also can play an important role. 
In the case in which the gas entropy enhancement 
of the southern subcluster had been  
caused by a first violent  encounter with the northern
subcluster, the cool core ($kT \sim 0.9$
keV -- as presented in Section \ref{sec.SB}) in the central galaxy of G345N (ESO 187-G026) 
could be the gas that survived core passage after a collision with the southern subcluster.
However, such a high velocity encounter would have highly disturbed
the X-ray morphologies of both G345 subclusters, which is not observed. 

For comparison, we searched for other clusters that present a similar
high central gas entropy ($K_0 \geq 150 \rm ~keV ~cm^2$) 
and low temperature ($kT \leq 3.2$ keV)
in the work of \citet{2009Cavagnolo}, and found:
A160, A193, A400, A562, A2125, and ZWCL 1215. Each  
cluster shows evidence of recent merger activity. 
A160 ($z = 0.0447$) has two giant ellipticals in the core,
that probably are in the process of merging. 
A193 ($z = 0.0485$) has a central galaxy with a triple nucleus 
\citep[IC 1695,][]{2003Seigar}. Since the typical time scale for multiple nuclei to 
merge into a single one is on the order of a Gyr \citep{2003Seigar}, 
observing a triple nucleus is a strong indication of recent merger activity. A400 ($z =
0.024$) is a well studied system which presents indications of merger
activity \citep{1992Beers}.
A562 ($z = 0.11$) has a Wide Angle Tail 
\citep[WAT -- radio lobes which are bent due to ram pressure as the host 
galaxy moves through the intracluster gas,][]{2011Douglas}. This is also a strong indicator
of merger activity.   
{\em Chandra} observations, together with multiwavelength data,
indicate that the A2125 complex ($z = 0.2465$) is probably undergoing major mergers \citep{2004Wang}.
Finally, inspecting the VLA FIRST image of ZWCL 1215 ($z = 0.075$), we notice a disturbed radio morphology associated 
with a galaxy (4C+04.41) that is probably merging from the southwest. However
this is not strong evidence of a major merger. 

Five of the six clusters with high
central gas entropy and low temperature show strong indications of recent or ongoing
merger activity which is likely responsible for the enhancement of the 
central gas entropy through shocks. This may suggest that the high
central gas entropy of G345S also was 
caused by a recent merger. However, if this were a merger with the
northern subcluster, we would expect to observe an
increase in the central entropy of the northern subcluster, which is
not observed, therefore making this scenario unlikely.  
         
The core of G345S hosts five giant ellipticals (see Figure \ref{fig:xmm_esored}) (unlike G345N, which has a single
dominant giant elliptical), 
suggesting that we are witnessing the formation of
the southern subcluster as groups of galaxies are in the process of
merging. This scenario can explain the high central entropy of G345S, as the
merging of the groups heats the gas through shocks. 

As an alternative to multiple mergers of groups, less violent
and frequent energy deposition into the
ICM could also produce a significant entropy enhancement. 
Isobaric heating is a reasonable approximation for a less violent
energy deposition where the gas pressure remains roughly constant
while part of the energy is converted into $pdV$ work, causing expansion of
the gas, and part is
transferred to internal energy, heating the gas. For isobaric heating, the added heat $\delta Q$
is related to the entropy increase by 
\begin{eqnarray}
\delta Q &=& \frac{5 kT_{\rm f} m_{\rm heated}}{2 \mu m_{\rm H}} 
  \left[1-\left(\frac{K_{\rm i}}{K_{\rm f}}\right)^{3/5}\right] = 7.9 \times
  10^{61} {\rm erg} \times \nonumber \\
&&  \left(\frac{kT_{\rm f}}{1 {\rm keV}}\right)
\left(\frac{0.6}{\mu}\right) 
\left(\frac{m_{\rm heated}}{10^{13} M_\odot}\right) 
 \left[1-\left(\frac{K_{\rm i}}{K_{\rm f}}\right)^{3/5}\right],\label{eq:entropy_increase}
\end{eqnarray} 
where $m_{\rm heated}$ is the gas mass that has been
heated (for isobaric heating, the heat increment equals the
enthalpy increment and $T_{\rm i}/T_{\rm f} = (K_{\rm i} / K_{\rm
  f})^{3/5}$, giving the equation above).

A less violent and steady source of energy that could explain the high central entropy 
could be energy injection into the ICM from an internal source such as
an AGN. 
We can estimate a global entropy by weighting the entropy by gas density: 
\begin{equation}
<K> = \frac{\int K \rho_{\rm g} ~ dV}{\int \rho_{\rm g} ~dV}.
\end{equation}

Within $r_{500} \sim$ 800 kpc the ratio of $<K>$ between G345S and G345N is $\sim$ 1.8.
From Equation (\ref{eq:entropy_increase}), we compute that 
the added heat necessary to increase the entropy of G345S is $\sim 
1.0 \times 10^{62} \rm ~erg$, which 
corresponds to an AGN power of $\sim 3.3 \times
10^{45}$--$10^{46} \rm ~erg~s^{-1}$ if all this energy 
has been injected into the ICM within 1 -- 0.1 Gyr, respectively. 
The 843 MHz radio image from the 
SUMSS survey \citep{1999Bock} of the G345 field shows a bright radio source associated with
one of the elliptical galaxies (ESO 187-IG 025 NED05 in G345S). Its radio power is 
$P_{\rm 1.4 ~GHz} = 1.4 \times 10^{24} \rm ~W~Hz^{-1}$, which
corresponds to a cavity power of roughly $10^{43}$--$\times 10^{46} \rm
~erg~s^{-1}$ \citep{2008Birzan,2011OSul}. Such an
AGN would need to inject energy into the ICM for 
a couple of Gyrs to enhance the central gas entropy by a factor of 1.8,
which also makes this scenario possible. Thus, either an AGN
sustained over a few Gyrs 
or the merger of groups could increase the central entropy of
G345S.

Departure from the scaling relation $K/K_{500} = 1.42 (r/r_{500})^{1.1}$  \citep{2010Pratt} 
is indicative of non-gravitational processes, where $K_{500}$ is
computed by:
\begin{eqnarray}
K_{500} &=& 106 ~ {\rm keV cm^2} \left( \frac{M_{500}}{10^{14} h_{70}^{-1}
M_\odot} \right)^{2/3} \left( \frac{1}{f_{\rm b}} \right)^{2/3}
\\ \nonumber
&\times& E(z)^{-2/3} h_{70}^{-4/3}, 
\end{eqnarray}
where $f_{\rm b} = 0.15$ is the baryon fraction \citep[this is the assumed
value for the baryon fraction in the work of][]{2010Pratt}. 
The right panel of Figure \ref{fig:entropy_prof} shows the
dimensionless entropy profile of both G345 subclusters. We see
that the entropy index of G345N is very close to the scaling relation
$K/K_{500} = 1.42 (r/r_{500})^{1.1}$ around $r_{500}$. On the other
hand, G345S largely exceeds this value at $r_{500}$, suggesting
that non-gravitational processes are playing a significant role in the
thermodynamics of the ICM even at $r_{500}$, 
supporting the scenario that multiple mergers of
groups have boosted the entropy index of G345S through shocks.


\section{Planck Determined and Expected Sunyaev-Zel'dovich Signals}

In this section, we compare the measured and expected
Sunyaev-Zel'dovich signals from G345.

The total Sunyaev-Zel'dovich signal is given by the integral of the Compton
parameter, $Y = \int y ~d\Omega$, where $\Omega$ is the solid
angle, and the Compton parameter $y$ is given by:
\begin{eqnarray}\label{eq:ysz}
  y = \frac{\sigma_{\rm T}}{m_{\rm e}c^2} \int_l kT_{\rm e}(r)n_{\rm
    e}(r) ~dl,
\end{eqnarray}
where $\sigma_{\rm T}$ is the Thomson cross section, $m_{\rm e}c^2$ is
the electron rest mass energy, $l$ is the distance along the 
line of sight, and $k$ is the Boltzmann constant. The total
Sunyaev-Zel'dovich signal also can be expressed as:
\begin{eqnarray}\label{eq:ysz}
  Y = \frac{\sigma_{\rm T}}{D_{\rm A}^2m_{\rm e}c^2} \int_V P ~dV,
\end{eqnarray}
where $D_{\rm A}$ is the angular size distance of the cluster, and $P$
is the electron pressure, $P = n_{\rm e} kT_{\rm e}$.
The spherical\footnote{The Planck Collaboration performs the integral
within a sphere, instead of performing it along the line of sight
to infinity (cylindrical integral).} Sunyaev-Zel'dovich signal measured by the {\em Planck} mission is
$Y = 0.0109 \pm 0.0032 \rm ~arcmin^2$ within an
angular size of $\theta = 5 \times \theta_{500} = 118.59 \rm ~arcmin$, where
$\theta_{500}$ is the angular size corresponding to
$r_{500}$ \citep{2011PlanckCol}. $\theta$ corresponds to 6.3 Mpc at the cluster redshift, which
leads to $r_{500} = 1260$ kpc, and 
$M_{500} = 5.94 \times 10^{14} M_\odot$ for the entire cluster. Based on the X-ray data, we
compute  $r_{500} \sim$ 700--800 kpc and 
$M_{500} \sim 1$--$2 \times 10^{14} M_\odot$ for each subcluster.

\begin{figure}[hbt!]
\centerline{%
\includegraphics[width=0.5\textwidth]{%
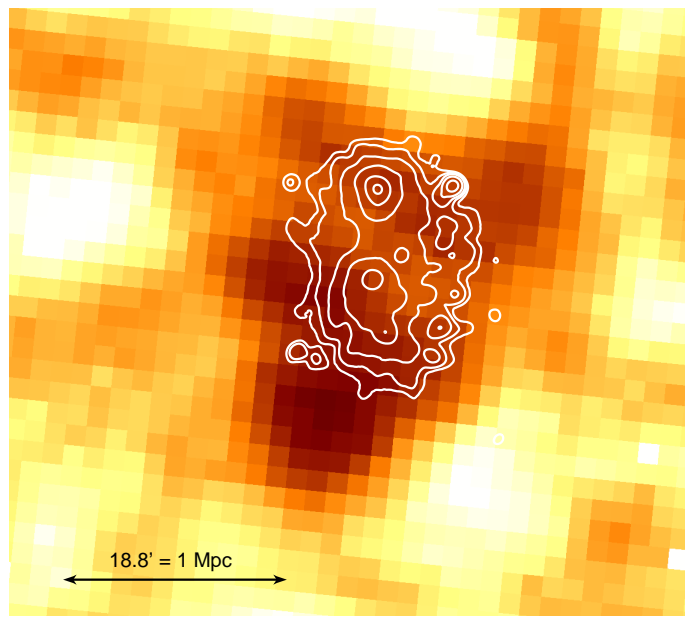}
}
\caption{
\small{
{\em Planck}-reconstructed $Y_{SZ}$ map of G345, overlaid with the
XMM-{\em Newton} X-ray isointensity
contours. Although we see a clear offset of $\sim 10'$ between the X-ray and the
Sunyaev-Zel'dovich signal peaks, this is consistent with {\em Planck's} much
courser spatial resolution and noise, compared to {\em Chandra}.
}
}
\label{fig:ysz_map_planck}
\end{figure}

Figure \ref{fig:ysz_map_planck} shows the {\em Planck}-reconstructed
$Y_{SZ}$ map of G345, overlaid with the X-ray isointensity
contours. Although we see a clear off-set between the X-ray and the
Sunyaev-Zel'dovich signal peaks, this is consistent with {\em Planck's} much
courser spatial resolution and higher instrumental noise compared to {\em Chandra}.
\citet{2013PlanckIntermediateVI} showed that an off-set as large as
5$'$ between the X-ray and Sunyaev-Zel'dovich signal peaks can be expected
in the reconstructed $y$-map for low significance
objects, due to astrophysical contributions and noise
fluctuations. They also showed that the SZ signal can be better reconstructed
assuming priors from other wavelengths, such as position, relative intensity between the
subclusters, and size.

\begin{figure}[hbt!]
\centerline{%
\includegraphics[width=0.5\textwidth]{%
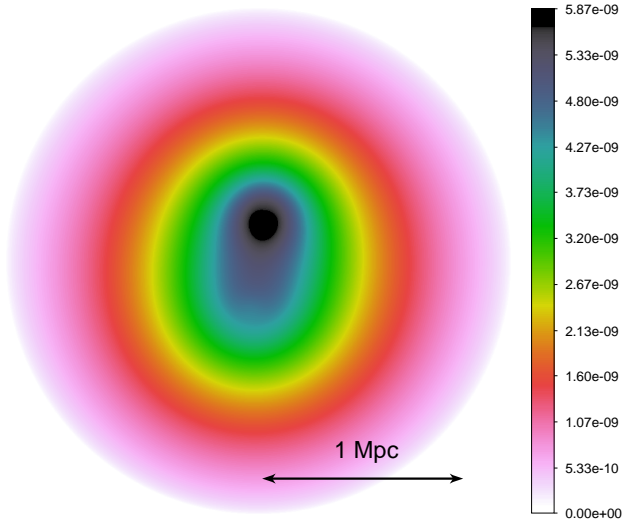}
}
\caption{
\small{
2.52 Mpc $\times$ 2.52 Mpc image of the spherically integrated Compton parameter $Y$
with 1 kpc $\times$ 1 kpc resolution. The values are in units of
arcmin$^2$. The centers of the
subclusters were separated by 400 kpc on the sky to match the X-ray observations. We see that most of the Sunyaev-Zel'dovich signal comes
from the northern component, since its gas density (and therefore pressure) is considerably
higher than that of G345S. 
}
}
\label{fig:ysz_map}
\end{figure}

To estimate the expected integrated Compton parameter, we used the electron density and
temperature profiles we determined for both subclusters to estimate the electron
pressure which we integrate along the line of sight, 
$P = n_{\rm e,N} kT_{\rm e,N} + n_{\rm e,S} kT_{\rm e,S}$, where the
indices N and S correspond to G345N and G345S. The centers of the
subclusters are separated by 400 kpc on the sky to match the  X-ray observations.
We computed the integrated Compton parameter
over a sphere of 1260 kpc (the $r_{500}$ given by the Planck
Collaboration) centered between the two subclusters (the resulting
2D map is presented in Figure \ref{fig:ysz_map}). The
\citet{2011PlanckCol}
determines the integrated Compton parameter within $5r_{500}$ to be $Y_{5R_{500}} =
1.81 Y_{500}$, where  $Y_{500}$ is the integrated Compton parameter
within $r_{500}$.  Using the X-ray derived parameters for the gas
temperature and density, we computed a $Y$
Compton parameter of  
$Y = 0.0136 \pm 0.0011 \rm ~ arcmin^2$, which is consistent within
$\sim ~0.8 \sigma$ of the
Planck Collaboration measured value of $Y = 0.0109 \pm 0.0032 \rm
~arcmin^2$. 

We also modified the pressure model from the universal profile
\citep{2010Arnaud} that was used to compute $Y$ by the Planck
Collaboration to a two component model that keeps the 
relative amplitude of the two components fixed to our X-ray model.
Using the Matched Multi Filters extraction algorithm,  
we obtained $Y = 0.0119 \pm 0.0025 \rm ~ arcmin^2$, which is
consistent within
$\sim ~0.6 \sigma$ of our X-ray measured value. This value is also
consistent with the previous Sunyaev-Zel'dovich signal measured
assuming a universal pressure profile. 

\section{A dynamical model for the G345 System}

In this section we apply the dynamical model presented by
\citet{1982Beers} to the G345N and G345S system to evaluate the dynamical state 
of the subclusters. 

For the case where the subclusters are gravitationally bound, 
we write the equations of motion in the following
parametric form:
\begin{eqnarray}
&&R=\frac{R_{\rm m}}{2}(1 - \cos \chi), \\
&&t=\left(\frac{R_{\rm m}^3}{8GM}\right)^{1/2}(\chi - \sin\chi), \\
&&V=\left(\frac{2GM}{R_{\rm m}}\right)^{1/2}\frac{\sin\chi}{(1 -
  \cos\chi)},
\end{eqnarray}
where $R_{\rm m}$ is the separation of the subclusters at maximum
expansion, $M$ is the total mass of the system, and $\chi$ is the 
development angle used to parametrize the equations. For the 
case where the subclusters are not gravitationally bound, 
the parametric equations are:
\begin{eqnarray}
&&R=\frac{GM}{V_\infty^2}(\cosh\chi - 1), \\
&&t=\frac{GM}{V_\infty^3}(\sinh\chi - \chi), \\
&&V=V_\infty\frac{\sinh\chi}{(\cosh\chi - 1)},
\end{eqnarray}
where $V_\infty$ is the asymptotic expansion velocity. The radial
velocity difference, $V_r$, and the projected distance, $R_p$,
are related to the system parameters by
\begin{eqnarray}
V_{\rm r} = V~\sin\alpha, ~ R_{\rm p} = R~\cos\alpha.
\end{eqnarray}

The total mass of the system is $M = (4.03 \pm 0.33) \times 10^{14}
~M_\odot$ (sum of the masses of both subclusters within $r_{200}$).  
We assume that the subclusters' velocities are the line of sight velocities of
their central dominant galaxies. We take $R_{\rm p}$ = 0.4 Mpc, the projected distance on the plane of
the sky between the dominant galaxies of each subcluster (see Figure \ref{fig:xmm_esored}). The redshift
difference between these galaxies yields a radial velocity difference of 
$V_r = 1134 \pm 66 {\rm~km~s^{-1}}$ \citep[the giant elliptical in
G345N is ESO 187-G026, at $z = 0.047176 \pm 0.000143$, and in G345S the
giant elliptical is ESO 187-IG025 NED04, at $z = 0.043223 \pm
0.000180$,][]{2004Smith}. 
We close the system of equations by setting $t$ = 12.86 Gyr, the age of the
Universe at the redshift of these clusters ($z=0.045$). 
These equations are then solved via an iterative procedure, 
which determines the radial velocity difference $V_{\rm r}$ as a
function of the projection angle $\alpha$.

Simple energy considerations specify the limits of the bound
solutions:
\begin{eqnarray}
V_{\rm r}^2  R_{\rm p} \leq 2GM~{\sin^2\alpha \cos\alpha}.
\label{eq:limit_bound_solutions}
\end{eqnarray}

\begin{figure}[hbt!]
\centerline{\includegraphics[width=.5\textwidth]{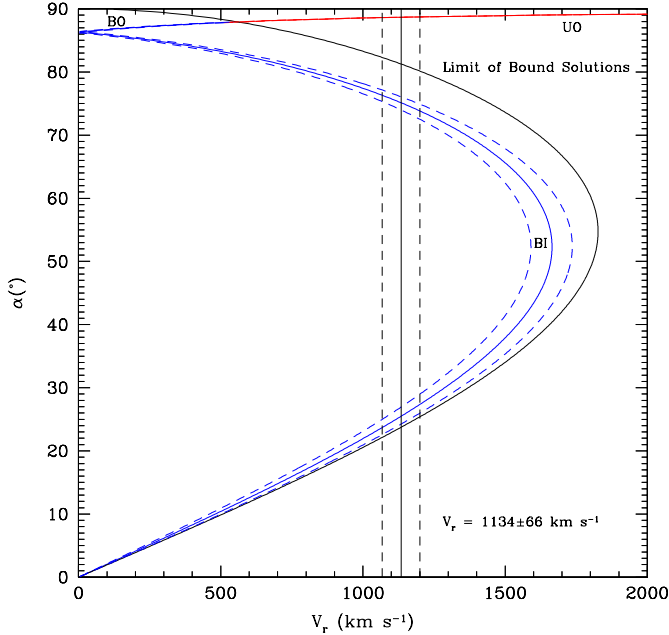}}
\caption{Projection angle $\alpha$ as a function of the radial
velocity difference $V_{\rm r}$ between the subclusters. BO, BI, and UO stand for Bound Outgoing, Bound
Incoming, and Unbound Outgoing solutions, respectively.
Solid blue and red lines correspond to bound and unbound solutions,
respectively. The vertical solid line corresponds to the radial velocity
difference between the giant ellipticals. Dashed lines correspond to
68$\%$ confidence ranges.
\label{fig:alpha_vr}}
\end{figure}

Figure \ref{fig:alpha_vr} shows the projection angle $\alpha$ as a function of the radial
velocity difference $V_{\rm r}$ between the subclusters. 
The uncertainties in the measured line-of-sight velocity and total
mass of the system lead to a range in the solutions for the
inclination angles ($\alpha_{\rm inf}$ and $\alpha_{\rm sup}$). 
The relative probabilities of these solutions are then computed by:
\begin{eqnarray}
p_i = \int_{\alpha_{{\rm inf},i}}^{\alpha_{{\rm sup},i}} \cos\alpha ~d\alpha,
\end{eqnarray}
where the index $i$ represents each solution. Finally, the probabilities can be
normalized by $P_{i} = p_{i}/(\sum_i p_i)$.

The radial velocity difference of the subclusters $V_r = 1134 \pm 66
{\rm~km~s^{-1}}$ yields two bound solutions and one unbound solution for $\alpha$. 
For the bound solutions, the subclusters are either approaching each other at
1173 km s$^{-1}$ (17$\%$ probability) or at 2636 km s$^{-1}$
(83$\%$ probability). The former solution corresponds to an encounter in less than 1.2 Gyr, given their 
separation of $\sim$ 1.56 Mpc. The latter corresponds to an encounter in less than 170 Myr, given their separation 
of $\sim$  440 kpc. The unbound solution (0.05$\%$ probability) corresponds to a separation of $\sim$ 17 Mpc.
These solutions are presented in Tables \ref{tab:bound} and
\ref{tab:unbound}. Given its low probability, the unbound solution can
be neglected, while the bound solution in which the separation
between the clusters is $\sim$ 440 kpc is highly favored ($83\%$
probability).

We note, that the dynamical analysis method from
\citet{1982Beers} assumes a purely radial infall (no
angular momentum). Also, the way the probabilities are computed 
favors small angle solutions. The small angle bound solution ($25.48^\circ$)
is highly favored ($P = 83\%$) compared to the other bound
solution ($75.13^\circ$ -- $P = 17\%$), despite its
supersonic infalling velocity ($\sim 2600 \rm ~km~s^{-1}$ -- the sound
speed of a 3 keV cluster is $\sim 900 \rm ~km~s^{-1}$, therefore $2600 \rm ~km~s^{-1}$
corresponds to a Mach number of $\sim$ 3). The virial
radius of these subclusters is roughly 1.2 Mpc. If, indeed, they were
separated by only $\sim$ 440 kpc, moving at $\sim 2600 \rm ~km~s^{-1}$,
shock discontinuities in the X-ray surface brightness would be seen 
in the region between their cores. Furthermore, fitting the
spectrum in a rectangular region (105 kpc $\times$ 345 kpc) between the subclusters gives 
${\rm k}T = 3.30_{-0.53}^{+0.84}$ at 68\% confidence, providing no evidence of shock heated gas
between the subclusters. In this analysis, the probabilities for the bound solutions should be
treated with caution, as we have no information about the angular momentum
of this system. 

\begin{deluxetable}{cccccc}
\tablecaption{Best Fit Parameters for the Bound Incoming Solutions of the Dynamical Model.} 
\tablewidth{0pt} 
\tablehead{ 
\colhead{$\chi$}&
\colhead{$\alpha$} &
\colhead{$R$} &
\colhead{$R_{\rm m}$} &
\colhead{$V$} &
\colhead{$P$} \\
\colhead{(rad)}&
\colhead{(degrees)} &
\colhead{(kpc)} &
\colhead{(kpc)} &
\colhead{($\rm km~s^{-1}$)} &
\colhead{($\%$)}
}
\startdata 
\chiba & \alphaba & \Rba & \Rmba & \Vba & 17 \\ 
\chibb & \alphabb & \Rbb & \Rmbb & \Vbb & 83    
\enddata
\tablecomments{Columns list best fit for the $\chi$ and $\alpha$
for the bound solutions of the dynamical model, and the corresponding values for
$R$, $R_{\rm m}$, $V$, and probability of each solution.}
\label{tab:bound}
\end{deluxetable}

\begin{deluxetable}{cccccc}
\tablecaption{Best Fit Parameters for the Unbound Outgoing Solution of the Dynamical Model.} 
\tablewidth{0pt} 
\tablehead{ 
\colhead{$\chi$}&
\colhead{$\alpha$} &
\colhead{$R$} &
\colhead{$V$} &
\colhead{$V_\infty$} &
\colhead{$P$} \\
\colhead{(rad)}&
\colhead{(degrees)} &
\colhead{(kpc)} &
\colhead{($\rm km~s^{-1}$)} &
\colhead{($\rm km~s^{-1}$)} &
\colhead{($\%$)}
}
\startdata 
\cchi & \aalpha & \R & \V & \Vinf & 0.05
\enddata
\tablecomments{Columns list best fit for the $\chi$ and $\alpha$
for the unbound solution of the dynamical model, and the corresponding values for
$R$, $V$, $V_\infty$, and probability of this solution.}
\label{tab:unbound}
\end{deluxetable} 

\newpage

\subsection{A Modified Dynamical Model}

Since the \citet{1982Beers} dynamical model is based on the timing
argument \citep[see e.g.~][]{1959Kahn}, which \citet{2008Li} have
shown to be biased and  over-constrained (failing to reproduce the
scatter observed in N-body simulations), we also investigate G345
using a modified version of the \citet{2013Dawson} Monte Carlo dynamic
analysis method\footnote{Monte Carlo Merger Analysis Code \citep[MCMAC][]{2014Dawson}.}.
The \citet{2013Dawson} method relaxes many of the constraints in the
timing argument and examines all merger scenarios consistent with the
observed state, enabling it to capture the observed scatter in the N-body simulations.
Additionally this method models the two subclusters as NFW halos
\citep{1996Navarro} so we can more accurately\footnote{To within
  $\sim$5-10\% agreement with N-body simulations \citep{2013Dawson}.}
estimate the time-till-collision ($TTC$), the period between collisions ($T$),
and the eventual relative collision velocity ($V_{\rm 3D}(t_{\rm
  col})$) \citep[for a precise definition of these quantities, please refer to][]{2013Dawson}.
We modified the \citet{2013Dawson} method slightly, since it is designed for post-merger systems and there is strong evidence that G345 is a pre-merger system.
We removed the prior that the time-since-collision\footnote{We define the time of collision to be the time of the first pericentric passage.} ($TSC$) be less than the age of the Universe, since $TSC$ is now recast as the $TTC$.
We also now allow for unbound scenarios, see Equation
(\ref{eq:limit_bound_solutions}), however we require that $V_{\rm r}$ be
less than the line-of-sight Hubble flow velocity for the unbound
realization to be considered valid,
\begin{equation}
V_{\rm r} \leq V_{\rm r, Hubble} = H(\bar{z})R\sin(\alpha),
\end{equation}
where $H(\bar{z})$ is the Hubble parameter at the average redshift of the two subclusters.
We summarize the results of this analysis in Table \ref{table:DawsonDynamics}. 

\begin{figure}[hbt!]
\centerline{
\includegraphics[width=0.45\textwidth,bbllx=98,bblly=191,bburx=480,bbury=584]{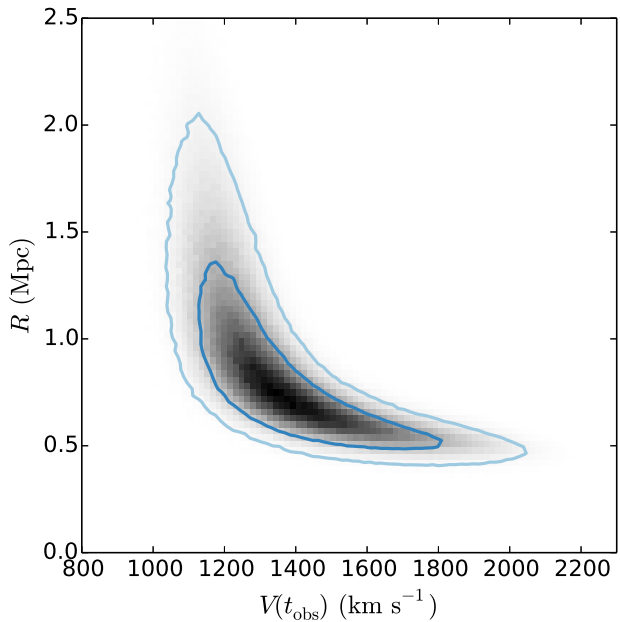}}
\caption{
The marginal posterior of the three-dimensional subcluster separation $d_{\rm 3D}$ and relative velocity $v_{\rm 3D}$ in the observed state $t_{\rm obs}$, inferred from the modified Dawson (2013) dynamics analysis. 
Dark and light blue contours represent the 68\% and 95\% confidence regions, respectively.
\label{fig:d3d-v3dobs}}
\end{figure}

\begin{figure}[hbt!]
\centerline{
\includegraphics[width=0.45\textwidth,bbllx=98,bblly=191,bburx=480,bbury=596]{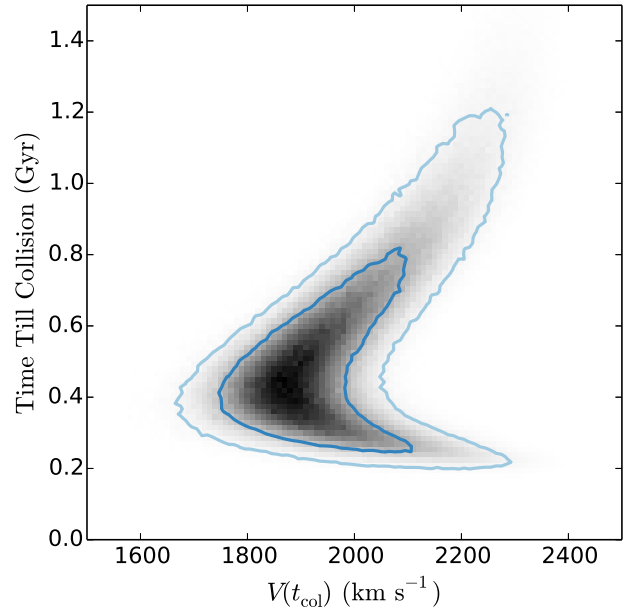}}
\caption{
The marginal posterior of the relative three-dimensional subcluster
velocity at the time of collision $v_{\rm 3D}(t_{\rm col})$ and the time it will take to collide given their observed state, inferred from the modified Dawson (2013) dynamics analysis. 
Dark and light blue contours represent the 68\% and 95\% confidence regions, respectively.
\label{fig:v3dcol-TTC}}
\end{figure}

From Figure \ref{fig:d3d-v3dobs} it can be seen that the large
relative velocity and small separation highly favored by the
\citet{1982Beers} dynamic model are disfavored by the modified
\citet{2013Dawson} dynamic model, as well as being inconsistent with the
observed gas properties.
However, there are still a number of realizations in the Monte Carlo
analysis where effects of the merger on the gas might be expected. In
principle these also could be excluded from the posterior distributions.
Based on this analysis, we find that the subclusters are likely to collide in 0.5
$\pm$ 0.2\,Gyr with a relative collision velocity of 2000 $\pm$
100\,km\,s$^{-1}$, see Figure \ref{fig:v3dcol-TTC}.
Based on this dynamic analysis, there is a slightly larger probability (2.6\% vs.~0.05\%) that the subclusters are unbound, but this scenario is still unlikely.
The unbound parameter estimates are also summarized in Table \ref{table:DawsonDynamics}.

\begin{deluxetable*}{lc|ccc|ccc}
\tablewidth{0pt}
\tablecaption{Parameter Estimates from Modified \citet{2013Dawson} Dynamics Analysis\label{table:DawsonDynamics}}
\tablehead{Parameter & Units  & \multicolumn{3}{|c|}{Bound Scenarios ($\mathcal{L}$=97.4\%)} & \multicolumn{3}{|c}{Unbound Scenarios ($\mathcal{L}$=2.6\%)} \\
     &   & \colhead{Location\tablenotemark{a}}  & \colhead{68\% LCL--UCL\tablenotemark{b}} & \colhead{95\% LCL--UCL\tablenotemark{b}}	& \colhead{Location\tablenotemark{a}}  & \colhead{68\% LCL--UCL\tablenotemark{b}} & \colhead{95\% LCL--UCL\tablenotemark{b}}	
}					
\startdata																							
M$_{\rm 200,N}$	&	$10^{14}$\,M$_\sun$	&	
1.6	&	1.4	--	1.8	&	1.1	--	2.0	&	
1.6	&	1.3	--	1.8	&	1.1	--	2.0	\\
M$_{\rm 200,S}$	&	$10^{14}$\,M$_\sun$	&	
2.5	&	2.3	--	2.8	&	2.0	--	3.0	&	
2.5	&	2.3	--	2.7	&	2.0	--	3.0	\\
$z_{\rm N}$	&		&	
0.0472	&	0.0470 --	0.0473	&	0.0469	--	0.0474	&	
0.0472	&	0.0470	--		0.0473	&	0.0469 --	0.0475	\\
$z_{\rm S}$	&		&	
0.0433	&	0.0431	--	0.0434	&	0.0429	--	0.0436	&	
0.0432	&	0.0430	--	0.0434	&	0.0429	--	0.0436	\\
$R_{\rm p}$	&	Mpc	&	
0.40	&	0.36	--	0.44	&	0.32	--	0.48	&	
0.41	&	0.37	--	0.45	&	0.33	--	0.50	\\
$\alpha$	&	degree	&	
59	&	46	--	70	&	37	--	77	&	
89	&	89	--	90	&	89	--	90	\\
$R$	&	Mpc	&	
0.81	&	0.60	--1.3	&	0.48	--2.0	&	
30	&	20	--	70	&	15	--	160	\\
$R_{\rm m}$	&	Mpc	&	
2.0	&	1.6	--	2.9	&	1.4	--5.7	&	
\nodata	&	\nodata	&	\nodata	\\
$V_{\rm 3D}(t_{\rm obs})$	&	km\,s$^{-1}$	&	
1300	&	1200	--	1600	&	1100	--	1900	&	
1100	&	1100	--	1200	&	1000	--	1300	\\
$V_{\rm 3D}(t_{\rm col})$	&	km\,s$^{-1}$	&	
2000	&	1900	--	2100	&	1800	--	2300	&	
\nodata	&	\nodata	&	\nodata	\\
$TTC$	&	Gyr	&	
0.5 	&	0.3	--	0.7	&	0.2	--	1.1	&	
\nodata	&	\nodata	&	\nodata	\\
$T$	&	Gyr	&	
4.9	&	3.7	--7.8	&	3.1	--	15	&	
\nodata	&	\nodata	&	\nodata	
\enddata														
\tablenotetext{a}{Biweight-statistic location \citep[see e.g.][]{1990Beers}.}						
\tablenotetext{b}{Bias-corrected lower and upper confidence limits, LCL and UCL respectively \citep[see e.g.][]{1990Beers}.}
\end{deluxetable*}

\section{Conclusions}

We have presented {\em Chandra}, XMM-{\em Newton}, and ROSAT observations of G345, 
a {\em Planck} detected  double cluster, and provided measurements of temperature, gas and total
masses, gas fraction, entropy profiles, expected Sunyaev-Zel'dovich
signal, and X-ray luminosities. 
Both the north and south subclusters have X-ray surface
brightnesses that are 
well described by a $\beta$-model within $r_{500}$, with $\beta \sim 0.5$. 
Both subclusters have gas masses in the range 1--2 $\times 10^{13} M_\odot$ and
total masses in the range 1--2 $\times 10^{14} M_\odot$, and gas mass
fractions in agreement, within the confidence range, of those found by \citet{2006Vik} for
clusters with similar total mass ($0.12 \pm 0.02$ for the northern 
subcluster and $0.09 \pm 0.01$ for the southern one).  
We show that the G345 
subclusters that 
are very likely ($> 97\%$ probability) gravitationally bound and
infalling and will collide in $500 \pm 200$ Gyr. 
We show that there is good ($0.8 \sigma$) agreement between the expected
Sunyaev-Zel'dovich signal predicted from the X-ray data and the measured value
from the {\em Planck} mission, and $0.6 \sigma$ agreement when the
{\em Planck} value is re-computed assuming a two component pressure
model, with relative amplitudes fixed using the X-ray results.
The high central entropy in G345S can be explained either by the
mergers of several groups, as suggested by the presence of five
massive elliptical galaxies or by AGN energy injection, as suggested
by the presence of a bright radio source in the massive elliptical
galaxy ESO 187-IG 025 NED05, or by a combination of both processes.

\acknowledgments

F.A-S. acknowledges support from {\em Chandra} grant G03-14131X. 
C.J., W.R.F. are supported by the Smithsonian Institution. 
R.J.W. is supported by NASA through the Einstein Postdoctoral grant
PF2-130104 awarded by the Chandra X-Ray Center, which is
operated by the Smithsonian Astrophysical Observatory for NASA under
contract NAS8-03060. P.E.J.N., A.V., L.P.D., and R.P.K. were supported
by NASA contract NAS8-03060.
Part of this work performed under the auspices
of the U.S. DOE by LLNL under Contract DE-AC52-07NA27344.
E.P., M.A. and G.W.P. acknowledge the support of the French
Agence Nationale de la Recherche under grant ANR-11-BD56-015.
We are also very grateful to the
anonymous referee who helped to improve this work.

\slugcomment{LLNL-JRNL-659077-DRAFT}

{\em Facilities:} CXO, XMM-{\em Newton}, ROSAT.

\appendix

\section{Stellar and Low Mass X-ray Binary Contributions to the X-ray Luminosity of Elliptical Galaxies}\label{appendix}

\subsection{Stellar Contribution}

\citet{2007Revnivtsev} showed that the unresolved X-ray halo in M32 can
be best explained by the cumulative emission from cataclysmic variables
and coronally active stars. In a following work,
\citet{2008Revnivtsev} used deep {\em Chandra} observations to measure
the unresolved X-ray emission in the  elliptical galaxy NGC 3379. They
suggested that the old stellar populations in all galaxies can be
described by a universal value of X-ray emissivity per unit stellar
mass or K-band luminosity. 

From the 2MASS K-band image of the giant elliptical in G345N (ESO 187-G026),
we extract its K-band luminosity and determine its stellar
mass. The K-band luminosity measured within 7 kpc from the center of
ESO 187-G026 (for comparison with the galaxy's X-ray luminosity measured within
the same region) is $L_{\rm K} = 1.6 \times 10^{43} \rm ~erg~s^{-1}$. 
\citet{2003Bell} showed that the relation between the K-band
luminosity and the stellar mass is given by:
\begin{eqnarray}
{\rm log}\left(\frac{M}{L_K}\right)=a_{\rm K}+b_{\rm K}\times (\rm color)\label{mass_Kluminosity}
\end{eqnarray}
where the $M/L$ ratio is given in solar units. For the $B-R$ color,
$a_{\rm K}$ = -0.264 and $b_{\rm K}$ = 0.138 \citep{2003Bell}. We calculate $B-R$ = 1.63
for ESO 187-G026, which gives $M = 1.78 \times 10^{11} M_\odot$.

The X-ray luminosity in the soft (0.5 -- 2.0 keV) band due to stellar emission
from ESO 187-G026 can now be computed using the
relation given by \citet{2007Revnivtsev}:

\begin{eqnarray}
L_{\rm 0.5-2.0~keV} = 7 \times 10^{38} \left(\frac{M}{10^{11}M_\odot}\right) \rm erg~s^{-1}\label{mass_Xluminosity}
\end{eqnarray}

Using Equation (\ref{mass_Xluminosity}) we compute a soft X-ray luminosity
of $L_{\rm 0.5-2.0~keV} = 1.25 \times 10^{39} \rm ~erg~s^{-1}$, $\sim$ 0.5 \%
of the total X-ray emission of the galaxy.

\subsection{Low Mass X-ray Binaries}

Low mass X-ray binaries (LMXB) also contribute to
the unresolved X-ray emission of galaxies. \citet{2004Gilfanov} showed that the 
distribution of near-infrared light in all galaxies closely traces the
azimuthally averaged spatial distribution of LXMBs. To describe it
quantitatively, they defined a template for the X-ray luminosity
function as a power law with two breaks:

\begin{eqnarray}
\frac{dN}{dL_{38}}=\left\{ \begin{array}{ll}
\renewcommand{\arraystretch}{3}
K_1 \left(L_{38}/L_{b,1}\right) ^{-\alpha_1},	
			& \mbox{\hspace{0.9cm} $L_{38}<L_{b,1}$}\\
K_2 \left(L_{38}/L_{b,2}\right)^{-\alpha_2},	
			& \mbox{$L_{b,1}<L_{38}<L_{b,2}$}\\
K_3 \left(L_{38}/L_{cut}\right)^{-\alpha_3},	
			& \mbox{$L_{b,2}<L_{38}<L_{\rm cut}$}\\
0,			& \mbox{\hspace{0.9cm} $L_{38}>L_{\rm cut}$}\\
\end{array}
\right.
\label{eq:uxlf}
\end{eqnarray}
where $L_{38}=L_X/10^{38} \rm~erg~s^{-1}$ and normalizations $K_1, K_2,$ and $K_3$ are
related by
\begin{eqnarray}
K_2=K_1 \left(L_{b,1}/L_{b,2}\right)^{\alpha_2}\nonumber \\
K_3=K_2 \left(L_{b,2}/L_{\rm cut}\right)^{\alpha_3}
\end{eqnarray}
They fixed the value of the high luminosity cut-off at
$L_{\rm cut}=500 \times 10^{38} \rm~erg~s^{-1}$. Due to the steep slope of the luminosity
function above $L_{b,2}$, the results are insensitive to the actual
value of $L_{\rm cut}$. Typically, for nearby galaxies, the source detection
threshold defines the luminosity cut-off. However, for 
ESO 187-G026 we cannot resolve any point sources, so the luminosity
cut-off is set to their fixed maximum value of $L_{\rm cut}=500 \times 10^{38} \rm~erg~s^{-1}$.

\citet{2004Gilfanov} provides the best fit value for the average
normalization of $K_1 = 440.4$ per $10^{11}~M_\odot$, and the
best fits to the parameters of the luminosity distribution of
$\alpha_1 = 1.0$, $\alpha_2 = 1.86$, $\alpha_3 = 4.8$, $L_{\rm b,1} = 0.19$,  
and $L_{\rm b,2} = 5.0$.

We can now compute the cumulative number of sources using:
\begin{eqnarray}
N_{\rm X}(>L) = \int_{L}^{L_{\rm cut}}\frac{dN}{dL_{38}}~dL_{38},
\end{eqnarray}
and the total luminosity by:
\begin{eqnarray}
L_{\rm X}(>L) = \int_{L}^{L_{\rm cut}}\frac{dN}{dL_{38}}~L_{38}~dL_{38},
\end{eqnarray}
which yields $L_{\rm 0.5-2.0~keV}(>10^{35}\rm ~erg~s^{-1}) = 1.62 \times 10^{40}
~erg~s^{-1}$ for ESO 187-G026. This leads to a (stellar + LMXB) luminosity of 
 $L_{\rm 0.5-2.0~keV} = 1.74 \times 10^{40} \rm~erg~s^{-1}$, 
which corresponds to $\sim 7\%$ of the galaxy's total 0.5--2.0 keV measured
X-ray luminosity within 7 kpc of its center.


\end{document}